\documentclass[12pt]{article}

\usepackage[dvipdfmx]{graphicx}
\usepackage{color}
\begin{document}
\newcommand{\NP}[1]{Nucl.\ Phys.\ {\bf #1}}
\newcommand{\PL}[1]{Phys.\ Lett.\ {\bf #1}}
\newcommand{\PR}[1]{Phys.\ Rev.\ {\bf #1}}
\newcommand{\PRL}[1]{Phys.\ Rev.\ Lett.\ {\bf #1}}
\newcommand{\PREP}[1]{Phys.\ Rep.\ {\bf #1}}
\newcommand{\PTP}[1]{Prog.\ Theor.\ Phys.\ {\bf #1}}
\newcommand{\PTEP}[1]{Prog.\ Theor.\ Exp.\ Phys.\ {\bf #1}}
\newcommand{\MPL}[1]{Mod.\ Phys.\ Lett.\ {\bf #1}}
\newcommand{\IJMP}[1]{Int.\ Jour.\ Mod.\ Phys.\ {\bf #1}}
\newcommand{\JHEP}[1]{JHEP\ {\bf #1}}
\begin{titlepage}
\setcounter{page}{0}
\begin{flushright}
\end{flushright}
~\\
\vspace{10mm}
\begin{center}
{\Large  CDT open-closed surface field theory of a 3D tensor-matrix model}

\vspace{10mm}

{\large Hiroshi\ Kawabe\footnote{e-mail address:
kawabe@yonago-k.ac.jp}} \\
{\em
National Institute of Technology, Yonago College\\ 
Yonago 683-8502, Japan} \\
\end{center}

\vspace{8mm}
\centerline{{\bf{Abstract}}}
\vspace{5mm}
We construct a tensor-matrix model which describes 3-dimensional (3D) Causal Dynamical Triangulation (CDT) of open-closed surface.
Though the usual splitting interaction of a surface is not derived from the stochastic quantization procedure, it provides another interaction of IK-type, which becomes the sole quantum correction.
Through the double scaling limit, it realizes CDT open-closed surface field theory including the IK-type interactions in the same condition with the closed surface CDT model.
Further, we investigate the commutation relations of the generators, for surfaces with the lowest numbers of boundary loops, in the Fokker-Planck (FP) Hamiltonian.
The generators seem to close in their commutation relations in our 3D model, differently from the 2D model, where the algebraic structure was not exact when commutators contain the generator of the open string edge.

\end{titlepage} 
\newpage
\renewcommand{\thefootnote}{\arabic{footnote}}
\setcounter{footnote}{0}
\section{Introduction}

Matrix models provide the formulation of dynamical triangulation (DT), or the random surface, on which discrete loops propagate and interact.
The stochastic quantization of the matrix models describes discrete loop evolution with the step of the stochastic time \cite{JR,Nak}.
Noncritical closed string field theories are realized through the double-scaling limit of these models, each of which possesses a string field Hamiltonian concerning the Virasoro algebra \cite{IK1,IK2}.
In addition to ordinary merging and splitting interactions, it is also possible to include an interaction extending the length of a string, simultaneously creating another string, like the pair creating counterpart of the extending part \cite{IK2}.
It is called as Ishibashi-Kawai (IK-) type interaction, which is originally an interaction of a loop in a spin cluster domain with the domain wall.
In the loop gas model, one loop field belongs to a domain $x$ of 1-dimensional (1D) discrete coordinate, instead of a spin domain, and it can interact with a loop in the neighboring domains $x\pm 1$ as well as the same domain $x$ \cite{Kos1,KK}.
The effective action of a matrix model for the loop gas model naturally includes the interacting term of two loops in the neighboring $x$ each other, or the origin of the IK-type interaction \cite{Kos2,Kos3}.
Noncritical open-closed string field theory is formulated by matrix-vector models, which possess an algebraic structure containing the Virasoro and current algebras \cite{Mog,AJ,NE,EKN}.
However, one of the problems of the DT models is that the possibility of splitting interaction becomes too large to realize stable propagation.
Causal dynamical triangulation (CDT) improves this situation because it is originally the model of only propagation with the time-foliation structure, in which any part of string propagates with the uniform pace.
While the causality forbids exactly both splitting and merging interactions, the model permitting only splitting interaction is found to be consistent as long as the baby strings eventually disappear without merging again into any string \cite{AL}.
This gentle breaking of the causality brings about a quantum effect to the CDT model, which is called generalized CDT (GCDT) \cite{ALWZ1}.
The string field model of GCDT is described as the merging coupling constant zero limit of the matrix model \cite{ALWWZ1,ALWWZ2,ALWWZ3}.
In these models, the stochastic time plays the role of time, or the geodesic distance of the world sheet \cite{ALWZ2}.
The GCDT model is able to include the IK-type interaction and this is also compatible with a matrix model \cite{FSW}. 
A natural development of these models is 3-dimensional (3D) CDT, in which a closed surface field evolves with a one-step propagator of shell composed of simplices like pyramids and tetrahedra \cite{AJL1}.
Numerical analyses of the 3D CDT clarify the possibility of the stable propagation depending on the coupling constants \cite{AJL2,AJLV,AJL3}.
Surface field theory is formulated by tensor models as the extension from the matrix models of 2-dimensional (2D) string field version \cite{Amb,BGR}.

In our previous works, we have constructed a novel type of GCDT matrix model, in which the effective action contains the one-step propagator of CDT \cite{Kaw1}.
As a variation of the loop gas model, we reinterpret the 1D discrete space-coordinate index of the original model as the discrete time, or the geodesic distance of CDT.
Then the stochastic time is not interpreted as the geodesic distance, but growth of interactions in the propagation.
Open-closed CDT string field theory describes the interactions of closed strings with D-branes through open strings.
It is formulated by a matrix-vector model, which is related to the same algebra as the DT model \cite{Kaw2}.
Furthermore, the closed surface field model of 3D CDT is constructed in the similar way by a tensor model \cite{Kaw3}.
The role of the IK-type interaction becomes more important as the sole quantum effect in the 3D surface model, because it is just another quantum effect as well as ordinary splitting interaction in the 2D string model, remaining in the continuum limit.

At this stage, the most interesting questions are as follows: 
whether the properties of 2D CDT of open-closed string are inherited in the 3D CDT models of open-closed surface and whether 3D CDT of closed surface is consistently extended to the non-trivial model of open-closed surface fields.
In this paper, we construct a tensor-matrix model which formulates 3D CDT open-closed surface field theory including the IK-type interactions.
The discrete surface, at every time, is constructed with squares related to tensors, and the boundary of open surface, or a closed loop, is expressed by connected links in the same manner as the matrix model.
Any propagator of a surface is formed by accumulating closed shells or open shells of the one-step propagators in the way of the time-foliation structure. 
While the one-step propagator of a surface is composed of pyramids and tetrahedra, the triangulation of a boundary loop propagator is embedded in the cross section at the boundary of an open shell.
We investigate the compatibility of the 3D surface model of tensor with the 2D string model of matrix.

The construction of this paper is as follows:
in section 2, we propose a tensor-matrix model of the 3D loop gas model.
Tensor fields express time-like triangles and space-like squares, whose edges correspond to the indices of tensors.
Matrices relate to time-like and space-like links that compose a triangulated one-step propagator of the boundary loop of an open surface.
Integrating out the time-like variables of triangles and links, we obtain an effective action with the invariant products of space-like variables of squares and links, expressing the one-step propagators of closed and open surfaces.
We formulate any propagator in finite steps of time with the effective action by the path integral procedure.
In section 3, we apply it the stochastic quantization method, which naturally derives the IK-type interactions in the same way as the 2D GCDT model.
In this model, the Fokker-Planck (FP) Hamiltonian describes the step of quantum process, but not the ordinary time evolution.
Through the double-scaling limit, in section 4, we estimate the possibility of realizing open-closed surface field theory with the IK-type interaction in keeping the CDT properties.
In section 5, we investigate the algebraic structure, or the commutation relations of generators, of which the FP Hamiltonian is composed.
The closure of generators in the commutation relation guarantees the consistency of the model, though it was not exact in the 2D open-closed string model because some commutators left terms with open string creation operators explicitly multiplied by generators.
The last section is devoted to the conclusions and discussions.
\section{CDT tensor-matrix model}

Two important properties of CDT are the causality and the time-foliation structure.
The tensor-matrix model, in the discrete level, may include some CDT breaking interactions which are expected to scale out in the continuum limit.
A closed surface, at any time $t$, is discretized with squares.
It is expressed as an invariant product of rank-four space-like tensors $(A_t)_{abcd}$, whose indices $a, b, c, d$ are assigned to the space-like links, or the sides of a square in turn.
A one-step propagator occupies the thin space between two surfaces of time $t$ and $t+1$, whose thickness is same everywhere over the shell according to the time-foliation structure.
The thickness equals to the length of the side of any triangle.
An up-triangle and a down-triangle, correspond to a rank-three time-like tensors $(B_t)_{aij}$ and $(C_t)_{aij}$, respectively.
The indices $i, j$ are assigned to the time-like links, or the sides of the triangles.
A square $(A_t)_{abcd}$ transfers to a site through an up-pyramid, ${\rm Tr}\{ A_t (B_t)^4 \}$, and vice versa through a down-pyramid, ${\rm Tr}\{ A_{t+1} (C_t)^4 \}$ (see Fig.\ref{fig:tensor}(I), (III)).
The connection between an up-pyramid and a down-pyramid is mediated by a tetrahedron, with two up-triangles and two down-triangles, expressed as a product ${\rm Tr}\{ (B_t)^2(C_t)^2 \}$ (see Fig.\ref{fig:tensor}(II)).
An open surface with a boundary is also made of squares.
The boundary possesses a sequence of indices which are still not contracted.
In order to make up an invariant of the open surface, we provide a matrix product, a closed chain of links, with tensor indices corresponding to the remaining ones on the boundary.
Each link relates to a space-like matrix field $\left( (\psi _t)_a \right)_{i_1 i_2}$ with a tensor index $a$, where the matrix indices $i_1, i_2$ are assigned to the sites on an orientable loop. 
The marginal cross-section of an open shell describes the one-step propagator of a boundary loop.
For this propagation, we need a time-like link $( (V_t)_i )_{i_1 i_2}$ to define two types of time-like triangle, an up-triangle ${\rm tr}\left( (\psi _t)_a (V_t)_i (V_t)_j \right)$ and a down-triangle ${\rm tr}\left( (\psi _{t+1})_a (V_t)_i (V_t)_j \right)$ (see Fig.\ref{fig:tensor}(IV), (V)).
The symbol ``tr'' means the invariant product with the contraction of the matrix indices of site, $i_1, i_2, \cdots$, while the capital ``Tr'' expresses the invariance with respect to the link index.
Taking product of a one-step boundary propagator of these triangles and the corresponding cross-section of an open shell, we obtain the invariant of the one-step propagator of an open surface (see Fig.\ref{fig:tensor}(VI)).
We assume that the indices $a,b,c,d,i,j,\cdots$ and $i_1,i_2,\cdots$ run from 1 to $N$.
The properties of fundamental tensor and matrix fields are as follows:
\begin{figure}[t]
\vspace{-35mm}
\hspace{0mm}
 \begin{minipage}{0.65\hsize}
  \begin{center}
   \scalebox{1}{\includegraphics[width=115mm, bb=0 0 960 720]{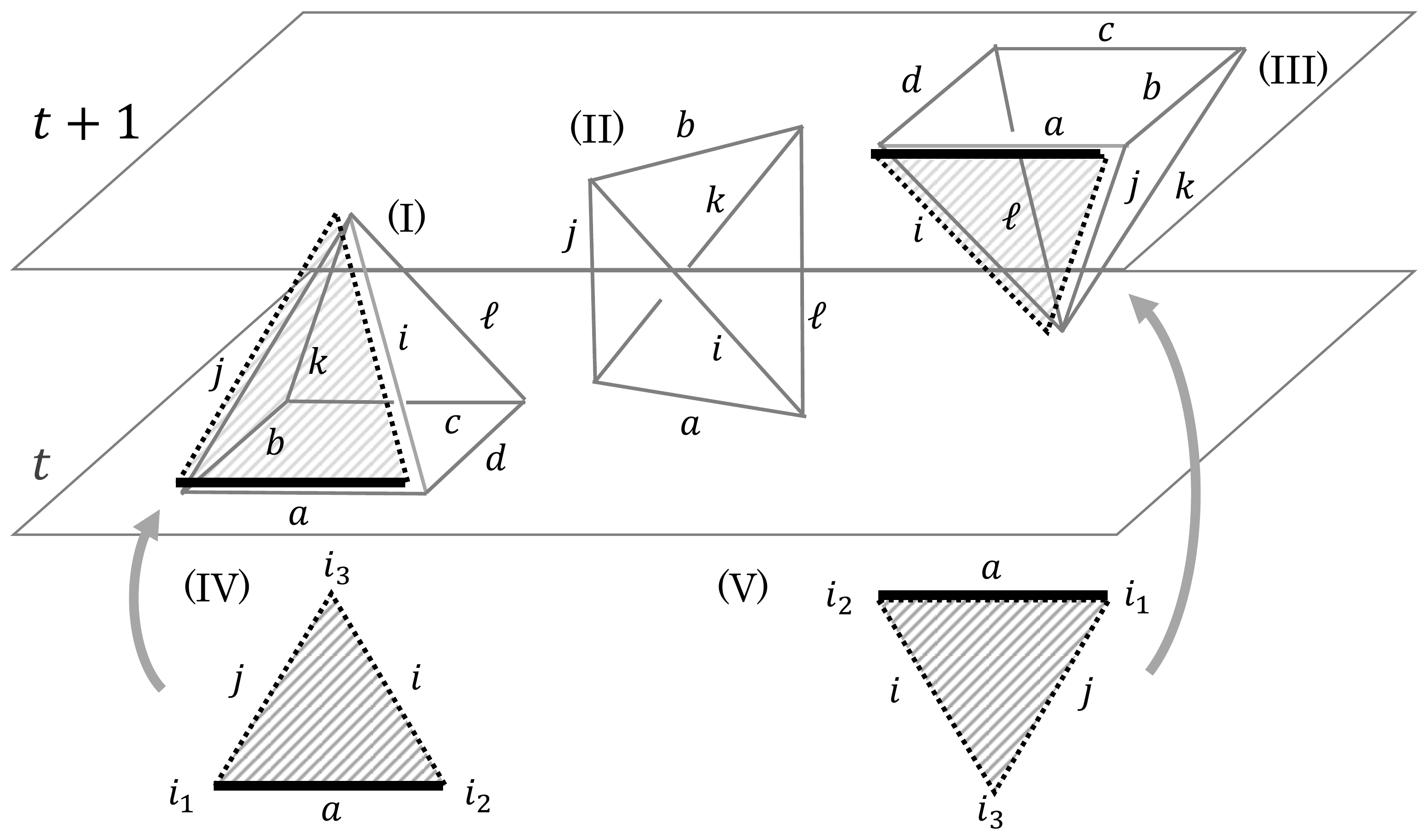}}
\end{center}
 \end{minipage}
\hspace{5mm}
 \begin{minipage}{0.35\hsize}
  \begin{center}
   \hspace{20mm}
   \scalebox{1}{\includegraphics[width=95mm, bb=00 0 960 720]{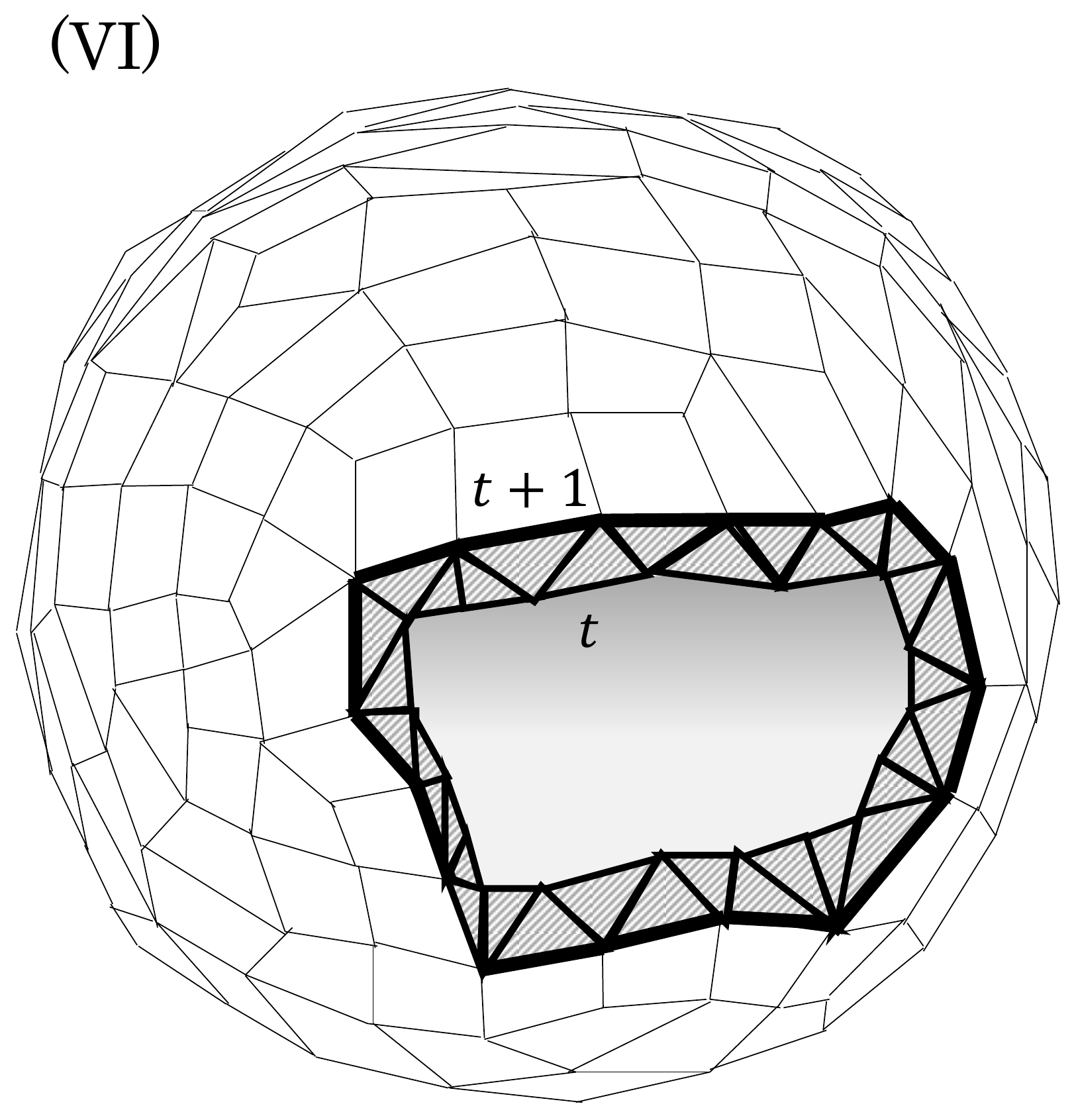}}
\end{center}
 \end{minipage}
\caption{
(I)~An up-pyramid ${\rm Tr}\! \left( A^*_{dcba} B_{aij} B_{bjk} B_{ck\ell} B_{d\ell i} \right)$, 
(II)~a tetrahedron ${\rm Tr}\! \left( B_{aij} B_{ak\ell} C_{bkj} C_{bi\ell} \right)$
and (III)~a down-pyramid ${\rm Tr} \left( A_{abcd} C_{aij} C_{bjk} C_{ck\ell} C_{d\ell i} \right)$ are three kinds of element of a one-step propagator.
For the open surface propagation, we need
(IV)~an up-triangle ${\rm tr} \left( (\psi_{t~a})_{i_1 i_2} (V_{t~i})_{i_2 i_3} (V_{t~j})_{i_3 i_1} \right)$, or the propagation of a boundary link with index $a$ to a site $i_3$, and 
(V)~a down-triangle ${\rm tr} \left( (\psi_{t+1~a})_{i_1 i_2} (V_{t~i})_{i_2 i_3} (V_{t~j})_{i_3 i_1} \right)$, or the propagation vice versa.
Triangles (IV) and (V) are attached to an up-triangle and a down-triangles, respectively, of simplices (I)$\sim$(III) at the boundary cross-section by the terms of the last line of the action (\ref{eq:action}), to construct the one-step propagator of the boundary of an open surface.
The shaded cross-section of (VI) is the one-step propagator of the boundary loop from time $t$ to $t+1$.
}
\label{fig:tensor}
\end{figure}
\begin{eqnarray}
&& (A_t)_{abcd}=(A_t)_{bcda}=(A_t)_{cdab}=(A_t)_{dabc},~~~~
 (A_t^*)_{abcd} \equiv (A_t)_{dcba}, \nonumber \\
&& (B_t^*)_{aij} \equiv (B_t)_{aji}, ~~~~ (C_t^*)_{aij} \equiv (C_t)_{aji}, \nonumber \\
&& \left( (\psi _t)_a \right)_{i_1 i_2}^* \equiv \left( (\psi _t)_a \right)_{i_2 i_1}, ~~~~\left( (V_t)_i \right) _{j_1 j_2}^* \equiv \left( (V_t)_i \right) _{j_2 j_1}, ~~~~{\rm tr}(*) \equiv (*)_{i_1 i_1}.
\end{eqnarray} 
We start with the action of the $U(N)_{\rm link}\times U(N)_{\rm site}$ gauge-invariant form:
\begin{eqnarray}
\label{eq:action}
S[A,B,C,\psi,V] & = & \sum_t {\rm Tr} \left[ -Ng (A_t)_{abba} + {1 \over 2} (A_t)_{abcd} (A_t)_{dcba} - {g \over 3N} (A_t)_{abdc} (A_t)_{cdfe} (A_t)_{efba} \right. \nonumber \\
 & & + {1 \over 2} \left\{ (B_t)_{aij} (B_t)_{aji} + (C_{t+1})_{aij} (C_{t+1})_{aji} \right\} \nonumber \\
 & & +{1 \over 2} \left\{ {\rm tr} \left( (\psi _t)_a  (\psi _t)_a \right) + {\rm tr} \left( (V _t)_i  (V _t)_i \right) \right\} \nonumber \\
 &  &  - {g^2 \over N^2} \left\{ (A^*_t)_{dcba} (B_t)_{aij} (B_t)_{bjk} (B_t)_{ck \ell} (B_t)_{d \ell i} \right. \nonumber \\
 &  & \left. ~~~~~~~+  (A_{t+1})_{abcd} (C_{t+1})_{aij} (C_{t+1})_{bjk} (C_{t+1})_{ck \ell } (C_{t+1})_{d \ell i} \right\} \nonumber \\
 &  & - {g \over N} (B_t)_{aij} (B_t)_{ak \ell} (C_{t+1})_{bi \ell} (C_{t+1})_{bkj} \nonumber \\
 &  & -{1 \over 2}{g_{\rm B} \over N} (A_t)_{ccba} {\rm tr} \left( (\psi _t)_a (\psi _t)_b \right) 
-{1 \over 4}{g_{\rm B} \over N^2} (A_t)_{dcba} {\rm tr} \left( (\psi _t)_a (\psi _t)_b (\psi _t)_c (\psi _t)_d \right)  \nonumber \\
 &  & \left. -{g_{\rm B} \over N} \left\{ (B_t)_{aij} {\rm tr} \left( (\psi _t)_a (V_t)_i (V_t)_j \right) + (C_{t+1})_{aij} {\rm tr} \left( (\psi _{t+1})_a (V_t)_i (V_t)_j \right) \right\} \right].
\end{eqnarray}
We have expressed only the index of link, but the index of site is omitted.
While each of the pyramids, ${\rm Tr}(AB^4)$ and ${\rm Tr}(AC^4)$, is multiplied by the factor $g^2 / N^2$, a tetrahedron, ${\rm Tr}(B^2 C^2)$, is weighted by $g/N$, corresponding to half the volume of a pyramid.
Five kinds of quadratic term connect the same types of face and link to build up a 3D space-time discretized propagator.
The invariants ${\rm Tr}(B^2)$ and ${\rm Tr}(C^2)$ glue time-like up-triangles and down-triangles, respectively, to connect simplices to form a shell.
The term ${\rm Tr}(A^2)$ glues space-like squares on every part of two one-step propagator shells neighboring each other; a square on the upside surface of the inner shell and another square on the downside surface of the outer shell.
The quadratic terms of small trace ${\rm tr}(\psi ^2)$ and ${\rm tr}(V^2)$ glue space-like and time-like links, respectively, of the boundary triangles to form the boundary loop propagator. 
Then, each term of the last line works to attach a boundary triangle, tr$(\psi V^2)$, to the corresponding triangle left unpaired at the boundary cross-section of an open shell, with the factor $g_{\rm B} / N$.
Here we complete the elements to express any one-step propagator as an invariant product for the open surface.

The linear term and the cubic term in the first line and two terms in the seventh line are not appropriate for CDT because they may break the time-foliation structure in the discrete level.
Although, they are found to be necessary for our model construction later.
Then, we define constructive fields of sphere $\phi _t (n)$ and closed surface with $h$ handles $\phi _t^{(h)}(n)$, composed of $n$ squares, or area $n$, by the invariant product, respectively,
\begin{eqnarray}
\label{eq:closed}
\phi _t (n) & \equiv & {\rm Tr} \left\{ \left( {A_t \over N} \right) ^n \right\} , \nonumber \\
\phi _t^{(h)} (n) & \equiv & {1 \over N^h} {\rm Tr} \left\{ \left( {A_t \over N} \right) ^n \right\},
~~~~\left( \phi _t^{(0)} (n) = \phi _t (n) \right).
\end{eqnarray}
An open surface of area $n$ with a boundary loop of length $k$ is defined as
\begin{eqnarray}
\label{eq:open1}
\omega _t (n|k) & \equiv & \left( {A_t \over N} \right) ^n_{a_k a_{k-1} \cdots a_2 a_1} 
\left( { (\psi _t)_{ a_1} \over \sqrt{N}} \right) _{i_1 i_2}
\left( { (\psi _t)_{a_2} \over \sqrt{N}} \right) _{i_2 i_3}
\cdot \cdot \cdot 
\left( { (\psi _t)_{a_{k-1}} \over \sqrt{N}} \right) _{i_{k-1} i_k}
\left( { (\psi _t)_{a_k} \over \sqrt{N}} \right) _{i_k i_1} \nonumber \\
& \equiv & \left( {A_t \over N} \right) ^n _{a_k a_{k-1} \cdots a_2 a_1} 
{\rm tr} \left( {\psi _t \over \sqrt{N}} \right)^k _{a_1 a_2 \cdots a_{k-1} a_k} \nonumber \\
& \equiv & {\rm Tr} \left\{ \left( {A_t \over N} \right) ^n {\rm tr} \left( {\psi _t \over \sqrt{N}} \right)^k \right\}.
\end{eqnarray}
When the open surface of area $n$ with a boundary of length $k$ has $h$ handles, we express it as
\begin{eqnarray}
\label{eq:open2}
\omega^{(h|1)} _t (n|k) & \equiv & {1 \over N^h} {\rm Tr} \left\{ \left( {A_t \over N} \right) ^n {\rm tr} \left( {\psi _t \over \sqrt{N}} \right)^k \right\}, ~~~~\left( \omega _t^{(0|1)}(n|k) = \omega _t (n|k) \right).
\end{eqnarray}
An open surface of area $n$ with no handle and two boundaries of lengths $k$ and $k'$ is written as
\begin{eqnarray}
\label{eq:open3}
\omega^{(0|2)} _t (n|k,k') & \equiv & 
 \left( {A_t \over N} \right) ^n_{a_k a_{k-1} \cdots a_2 a_1,~b_{k'} b_{k'-1} \cdots b_2 b_1} 
{\rm tr} \left( {\psi _t \over \sqrt{N}} \right)^k _{a_1 a_2 \cdots a_{k-1} a_k} 
{\rm tr} \left( {\psi _t \over \sqrt{N}} \right)^{k'} _{b_1 b_2 \cdots b_{k'-1} b_{k'}} \nonumber \\
& \equiv & {\rm Tr} \left\{ \left( {A_t \over N} \right) ^n {\rm tr} \left( {\psi _t \over \sqrt{N}} \right)^k {\rm tr} \left( {\psi _t \over \sqrt{N}} \right)^{k'} \right\}. 
\end{eqnarray}
The integral of the surface fields with the factor $e^{-S}$ over the tensor and matrix fields counts all the possible ways of discretized propagator.
Integrating out all time-like variables $B, C$ and $V$ in the partition function, we obtain the effective action $S_{\rm eff}$ through
\begin{eqnarray}
\label{eq:partition}
Z = \int {\cal D} A {\cal D} B {\cal D} C {\cal D} \psi {\cal D} V e^{-S[A,B,C,\psi ,V] } = \int {\cal D} A {\cal D} \psi e^{-S_{\rm eff}[A, \psi]} .
\end{eqnarray}
The effective action composed of only space-like tensor $A$ and matrix $\psi$ is,
\begin{eqnarray}
\label{eq:effective}
S_{\rm eff} & = & \sum_t \left[ (\rm{potential~terms}) \right. \nonumber \\
& & - \sum_{n,m} \sum_s g^{2n+2m+s} \sum_L C(n,m,s,L) N^{-n-m-s+L} {\rm Tr} \left\{ \left( {A_t \over N} \right)^n \right\} {\rm Tr} \left\{ \left( {A_{t+1} \over N} \right)^m \right\} \nonumber \\
& & - \sum_{n,m} \sum_{k,\ell} g_{\rm B}^{k+\ell} \sum_s g^{2n+2m+s} \sum_L O(n,m,s,L|k,\ell) N^{-n-m-s+L+{k+\ell \over 2}}  
\nonumber \\
& & \hspace{7em} \left. 
\times{\rm Tr} \left\{ \left( {A_t \over N} \right) ^n {\rm tr} \left( {\psi _t \over \sqrt{N}} \right)^k \right\}
{\rm Tr} \left\{ \left( {A_{t+1} \over N} \right) ^m {\rm tr} \left( {\psi _{t+1} \over \sqrt{N}} \right)^{\ell} \right\}
+\cdots \right] . 
\end{eqnarray}
The power indices of $g$ and $g_{\rm B}$ count the volume of a thin one-step propagating shell and the area of the cross section swept by a boundary loop, respectively.
Each term of the second line contains two factors corresponding to two closed surfaces in time $t$ and $t+1$, or an inside surface with $n$ squares and an outside one with $m$ squares, respectively.
The coefficient $C(n,m,s,L)$ is the configuration number of CDT discretization for the one-step propagating space, between the two surfaces, made of $n$ up-pyramids, $m$ down-pyramids and $s$ tetrahedra, containing $L$ inner links.
Here, $L$ is the number of the time-like links, along which the edges of simplices are attached from all directions.
Although it corresponds to the binomial coefficient in the 2D CDT model, we do not have enough knowledge in the 3D CDT model.
In the terms of the last two lines we have two factors of open surfaces in time $t$ (and $t+1$), with area $n$ ($m$) and the boundary loop length $k$ ($\ell$), respectively.
Another coefficient $O(n,m,s,L|k,\ell)$ counts the configuration of the propagating space of the open surface characterized by, in addition to $n,m,s,L$, the link numbers of boundary loops $k$ and $\ell$.
Analogous to the 2D CDT model, the one-step propagators of the closed surface and the open surface are defined by
\begin{eqnarray}
\label{eq:1step-propagator}
G^{(h)}(n,m) & \equiv & \sum_s g^{2n+2m+s} \sum_L C(n,m,s,L) \delta_{-n-m-s+L,2-2h} , \nonumber \\
F^{(h|b =1)}(n,m|k,\ell) & \equiv & g_{\rm B}^{k+\ell} \sum_s g^{2n+2m+s} \sum_L O(n,m,s,L|k,\ell) \delta_{-n-m-s+L+{k+\ell \over 2},2-2h-b} (b=1), \nonumber \\
~
\end{eqnarray}
respectively.
The power indices of $N$ satisfy $-n-m-s+L=2-2h$ for the one-step propagator of a closed surface with $h$ handles \cite{Kaw3}.
For an open surface, we conjecture a similar relation $-n-m-s+L+{k+\ell \over 2}=2-2h-b$, where $b$ is the number of the boundary loops. 
We can rewrite the effective action with the surface fields and the one-step propagators.
The effective action is the same form as 2D CDT open-closed string field theory and it is divided into two parts, the CDT exact part $S_0$ and the CDT breaking part $S_1$,
\begin{eqnarray}
\label{eq:effectiveaction}
S_{\rm eff} & = & S_0 + S_1 ,  \nonumber \\
S_0 & = & N^2 \sum_t \left[  {1 \over 2} {\rm Tr} \left\{ \left({A_t \over N}\right)^2 \right\} - \sum_h \sum_{n,m} \phi _t^{(h)}(n) G^{(h)}(n,m) \phi _{t+1}^{(h)}(m) \right. \nonumber \\
& & \left.  + {1 \over 2}{1 \over N} {\rm Tr} \left\{ {\rm tr} \left({\psi _t \over \sqrt{N}} \right)^2 \right\}  - {1 \over N} \sum_h \sum_{n,m} \sum_{k,\ell} \omega _t^{(h|1)}(n|k) F^{(h|1)}(n,m|k,\ell) \omega _{t+1}^{(h|1)}(m|\ell) 
+\cdots \right] , \nonumber \\
S_1 & = &  N^2 \sum_t \left[ -g {\rm Tr} \left\{ {A_t \over N}\right\} - {g \over 3} {\rm Tr} \left\{ \left({A_t \over N}\right)^3 \right\} \right. \nonumber \\
& & \left. -{1 \over 2} {g_{\rm B} \over N} {\rm Tr} \left\{ \left( {A_t \over N} \right) {\rm tr} \left({\psi _t \over \sqrt{N}}\right)^2 \right\} -{1 \over 4} {g_{\rm B} \over N} {\rm Tr} \left\{ \left({A_t \over N}\right){\rm tr} \left({\psi _t \over \sqrt{N}} \right)^4 \right\} \right] .
\end{eqnarray}
The dots contain terms of one-step propagators with the open surfaces which possess two or more boundary loops.
We write down them explicitly as,
\begin{eqnarray}
\label{eq:effectiveaction2}
&& \sum_{b=2}^{\infty} {1 \over N^{b}} \sum_h \sum_{n,m} \sum_{k_1,\ell_1} \cdots \sum_{k_{b},\ell_{b}} \omega _t^{(h|b)}(n|k_1,k_2,\cdots ,k_{b} ) \nonumber \\
&&\hspace{5em}\times F^{(h|b)}(n,m|k_1, k_2 \cdots , k_{b} ; \ell_1, \ell_2, \cdots ,\ell_{b}) \omega _{t+1}^{(h|b)}(m|\ell_1, \ell_2, \cdots ,\ell_{b}) .
\end{eqnarray}
Hereafter we omit these terms in the case without necessity.

When the surfaces at time $t$ and $t+1$ are spheres, $h=0$, the one-step propagator is a discretized spherical shell.
If we slice the shell at time $t+{1 \over 2}$, the cross section expresses the `quadrangulation' of the sphere with $n+m+s$ squares.
It has the same structure as the matrix model with the site index of the matrix field replacing the link index of the original tensor field.
The dominant quadrangulation, weighted with $N^2$, is the planar diagram, which corresponds to the discretization of 3D shell with pyramids and tetrahedra disposed in the planar way.
Complicated contractions of tensors, which we might not have corresponding manifold, are suppressed with the additional orders of $1/N$ relative to the planar contractions.
In the same way, the one-step propagations of torus (and closed surfaces with two or more handles) are dominated by the discretization whose corresponding $t+{1 \over 2}$ quadrangulation are torus (and closed surfaces with the same handle numbers), respectively.
In general, open-closed shells are weighted with $N^{2-2h-b}$ as the dominant order.
Our tensor model may derive constructive propagator with entangled contraction of tensors, in addition to the one whose corresponding quadrangulated cross section becomes the surface with the same topology.
To the latter we restrict as the contribution to the one-step propagators in Eq.(\ref{eq:effectiveaction}), while we omit the former with the different topology as higher order of $1/N$. 

In the above expression, we have defined the closed surface field $\phi _t(n)$ with respect only to the area $n$, ignoring the difference of the configuration of squares related to the ways of contracting tensors.
We will call the alternative field classified by the configurations of squares as `quadrangulated expression' and express with hat, $\hat{\phi}$ and $\hat{\omega}$.
Then our surface fields contain any hatted ones,
\begin{eqnarray}
\label{eq:expression1}
\phi _t^{(h)} (n) & = & \{ \hat{\phi}_t^{(h)}(n;\alpha) | \alpha \in {\rm any ~configuration ~with ~{\it n} ~squares} \}, \nonumber \\
\omega _t^{(h|1)} (n|k) & = &  \{ \hat{\omega}_t^{(h|1)}(n|k;\alpha) | \alpha \in {\rm any ~configuration ~with ~{\it n} ~squares  ~and ~{\it k} ~links} \},
\end{eqnarray}
where the additional index $\alpha$ is assigned to each individual configuration.
Accordingly, the configurations of one-step propagator are also classified by those of the squares in the initial and final times.
A one-step propagator in our usual expression, or `number expression', is the sum of the various hatted ones of the quadrangulated expression, with only the numbers of square and boundary link fixed;
\begin{eqnarray}
\label{eq:expression2}
G^{(h)}(n,m) & = & \sum_{\alpha, \beta} \hat{G}^{(h)}(n,m;\alpha,\beta)   , \nonumber \\
F^{(h|1)}(n,m|k,\ell) & = & \sum_{\alpha, \beta} \hat{F}^{(h|1)}(n,m|k,\ell;\alpha,\beta) . 
\end{eqnarray}
The one-step propagation terms in $S_0$ are understood exactly as
\begin{eqnarray}
\label{eq:expression3}
\phi _t^{(h)} (n) G^{(h)}(n,m) \phi _{t+1}^{(h)} (m)& \equiv & \sum_{\alpha, \beta} \hat{\phi}_t^{(h)}(n;\alpha) \hat{G}^{(h)}(n,m;\alpha,\beta) \hat{\phi}_{t+1}^{(h)}(m;\beta)  , \nonumber \\
\omega _t^{(h|1)} (n|k) F^{(h|1)}(n,m|k,\ell) \omega _{t+1}^{(h|1)} (m|\ell) & \equiv & \sum_{\alpha, \beta} \hat{\omega}_t^{(h|1)}(n|k;\alpha) \hat{F}^{(h|1)}(n,m|k,\ell;\alpha,\beta) \hat{\omega}_{t+1}^{(h|1)}(m|\ell;\beta). \nonumber \\
~
\end{eqnarray}
The quadrangulated expression is appropriate to realize the composition rule of the time-foliation structure.
The exact CDT propagator of a disc with boundary number $b =1$, from $\omega ^{(h|1)}_0 (n_0|k_0)$ to $\omega ^{(h|1)}_t (n_t|k_t)$ in the finite time $t$, is obtained utilizing Eq.(\ref{eq:expression3}) as
\begin{eqnarray}
\label{eq:expression4}
&&\langle \omega ^{(h|1)}_0 (n_0|k_0) \omega ^{(h|1)}_t (n_t|k_t) \rangle_0 \nonumber \\
&&= \sum_{\alpha _0} \sum_{\alpha _t} \langle \hat{\omega}^{(h|1)}_0 (n_0|k_0 ;\alpha_0) \hat{\omega}^{(h|1)}_t (n_t|k_t ;\alpha_t) \rangle_0 \nonumber \\
&& \equiv  \sum_{\alpha _0} \sum_{\alpha _t} {1 \over Z_0} \int {\cal D} A {\cal D}\psi  \hat{\omega}^{(h|1)}_0 (n_0|k_0 ;\alpha_0) \hat{\omega}^{(h|1)}_0 (n_t|k_t ;\alpha_t) e^{-S_0} \nonumber \\
& & =  N^{(2-2h-1)t} \sum_{n_1, \cdots , n_{t-1}=1} ^{\infty} \sum_{k_1, \cdots , k_{t-1}=1} ^{\infty} \sum_{\alpha_0, \cdots , \alpha_t } 
 k_t \prod_{i=0}^{t-1} k_i \hat{F}^{(h|1)} (n_i,n_{i+1}|k_i,k_{i+1};\alpha_i ,\alpha_{i+1}), 
\end{eqnarray}
where $Z_0$ is the partition function with only the CDT exact part $S_0$ of the effective action.
The same type of time-foliation structure is also possible for the closed surface propagator \cite{Kaw3}.
The CDT breaking part $S_1$ causes propagation of surfaces increasing and decreasing the area and the boundary length in the same time-slice, hence violates the time-foliation structure.
Although these interactions seem to disturb the construction of the CDT model, it will be found to be rather necessary in a scaling limit.
Because our interest is focused on the scaling behavior of each interaction in the continuum limit, not on the configuration, we need not use the quadrangulated expression.
Then we proceed in the number expression, where the fields are decided with the numbers of squares and boundary links.

\section{Stochastic quantization}

We apply the stochastic quantization method to the above tensor-matrix model to obtain the IK-type interactions.
The Langevin equations of the tensor field $A_t$ and the matrix field $\psi _t$ are derived from the effective action,
\begin{eqnarray}
\label{eq:langevinA}
\Delta (A_t)_{abcd} & = & - \Delta \tau {{\partial S_{\rm eff}} \over {\partial (A_t)_{dcba}}} + (\Delta \xi _t)_{abcd} ,
\end{eqnarray}
\begin{eqnarray}
\label{eq:langevinpsi}
\Delta \left(( \psi _t )_a \right)_{i_1 i_2} & = & -\lambda _{\rm B} \Delta \tau {\partial S_{\rm eff} \over \partial ((\psi _t)_a)_{i_2 i_1} } + \left( (\Delta \eta _t)_a \right)_{i_1 i_2} ,
\end{eqnarray}
respectively.
$\lambda _{\rm B}$ is the scale parameter of the stochastic time evolution on the boundary.
The last terms of Eqs.(\ref{eq:langevinA}) and (\ref{eq:langevinpsi}), the white noise terms, satisfy the following correlations:
\begin{eqnarray}
\label{eq:correlation}
\langle (\Delta \xi _t)_{abcd} (\Delta \xi _{t'})_{d'c'b'a'} \rangle _{\xi}  & = & 
{1 \over 2} \Delta \tau \delta _{tt'}  (\delta _{aa'} \delta _{bb'} \delta_{cc'} \delta_{dd'} 
+\delta _{ba'} \delta _{cb'} \delta_{dc'} \delta_{ad'} \nonumber \\
&&~~~~~~~~+\delta _{ca'} \delta _{db'} \delta_{ac'} \delta_{bd'} +\delta _{da'} \delta _{ab'} \delta_{bc'} \delta_{cd'}  ), \nonumber \\
\langle \left( (\Delta \eta _t)_a \right)_{i_1 i_2}  \left( (\Delta \eta _{t'})_{a'} \right)_{j_1 j_2}  \rangle _{\eta}  & = & 
2 \lambda _{\rm B} \Delta \tau \delta _{tt'}  \delta _{aa'} \delta _{i_1 j_2} \delta_{i_2 j_1} .
\end{eqnarray}
The stochastic time evolution of the closed surface variable is
\begin{eqnarray}
\label{eq:langevinphi}
\Delta \phi _t^{(h)} (n) &=& \Delta \tau n \left[ g \phi _t^{(h)} (n-1) - \phi _t^{(h)} (n) + g \phi _t^{(h)} (n+1)  + {1 \over N}(n-1) \phi _t^{(h+1)} ( n-2 ) \right. \nonumber \\
& & +{1 \over N} g_{\rm B} \omega_t^{(h|1)} (n-1|2)  +{1 \over N} g_{\rm B} \omega_t^{(h|1)} (n-1|4) \nonumber \\
 & & + \sum_{h'=0}^{\infty} \sum_{n'=1}^{\infty} \phi _t^{(h+h')} (n+n'-2) n' \sum_{m=0}^{\infty} \left\{ G^{(h')}(n',m) \phi _{t+1}^{(h')} (m) + G^{(h')}(m,n') \phi _{t-1}^{(h')} (m) \right\} \nonumber \\
& & + {1 \over N} \sum_{h'=0}^{\infty} \sum_{n',k} \omega _t^{(h+h'|1)} (n+n'-2|k) \nonumber \\
& & \hspace{20mm} \times n' \sum_{m, \ell} \left\{ F^{(h'|1)}(n',m|k,\ell) \omega _{t+1}^{(h'|1)} (m|\ell ) + F^{(h'|1)}(m,n'|\ell ,k) \omega _{t-1}^{(h'|1)} (m|\ell ) \right\} \nonumber \\
 & & \left.+ \cdots ~\right] + \Delta \zeta _t^{(h)} (n),
\end{eqnarray}
where the last term is the constructive noise term of the closed surface field defined by
\begin{eqnarray}
\label{eq:constructive1}
\Delta \zeta _t^{(h)} (n) & \equiv & {n \over N^{h+1}} \left( {A_t \over N} \right)^{n-1}_{dcba} (\Delta \xi_t )_{abcd}~~~~\left(\Delta \zeta _t (n) \equiv \Delta \zeta_t^{(0)} (n) \right).
\end{eqnarray}
The terms in the third line of Eq.(\ref{eq:langevinphi}) express the expanding of the original closed surface, like the instantaneous growth of a balloon from a point on the surface, simultaneously creating another sphere in either time $t+1$ or $t-1$.
This baby sphere with area $m$ in the neighboring time is related to the inflating balloon part of area $n'$ on the original surface through the one-step propagator.
This is the IK-type interaction, which is different from ordinary separation of one sphere into two pieces with preserving the total number of squares.
We have another IK-type interaction from the fourth to the fifth lines, concerning the pair-creation of open surfaces.
The original closed surface changes to an expanding open surface, as the sudden growth of a half-sphere at one point, with creating a partner open surface in either of neighboring times.

In the same way, we have the evolution of the open surface variable,
\begin{eqnarray}
\label{eq:langevinomega}
\Delta \omega _t^{(h|1)} (n|k) & = & \Delta \tau n \left[
g \omega_t^{(h|1)} (n-1|k) - \omega_t ^{(h|1)} (n|k) +g \omega_t^{(h|1)} (n+1|k) \right. \nonumber \\
& & + {1 \over N}(n-1) \omega_t^{(h+1|1)} (n-2|k) +{1 \over N} g_{\rm B} \omega_t^{(h|2)}(n-1|k,2) +{1 \over N} g_{\rm B} \omega_t^{(h|2)}(n-1|k,4) \nonumber \\
& & + \sum_{h'=0}^{\infty} \sum_{n'=1}^{\infty} \omega _t^{(h+h'|1)} (n+n'-2|k) \nonumber \\
& & \hspace{30mm} \times n' \sum_{m=0}^{\infty} \left\{ G^{(h')}(n',m) \phi _{t+1}^{(h')} (m) +G^{(h')}(m,n') \phi _{t-1}^{(h')} (m) \right\}\nonumber \\
& & + {1 \over N} \sum_{h'=0}^{\infty} \sum_{n',k'} \omega _t^{(h+h'|2)} (n+n'-2|k, k') \nonumber \\
& & \hspace{15mm} \times n' \sum_{m, \ell} \left\{ F^{(h'|1)}(n',m|k,\ell) \omega _{t+1}^{(h'|1)} (m|\ell ) + F^{(h'|1)}(m,n'|\ell ,k) \omega _{t-1}^{(h'|1)} (m|\ell ) \right\} \nonumber \\
& & \left.+ \cdots ~\right] + \Delta \zeta _t^{(h|1)} (\bar{n}|k) \nonumber \\
& & + \lambda _{\rm B} \Delta \tau k \left[ -\omega_t^{(h|1)} (n|k) + g_{\rm B} \omega_t^{(h|1)}(n+1|k) + g_{\rm B} \omega_t^{(h|1)}(n+1|k+2) \right. \nonumber \\
& & + {1 \over N} \sum_{k'=0}^{k-2} \omega_t^{(h|2)}(n|k',k-k'-2) \nonumber \\
& & + \sum_{h'=0}^{\infty} \sum_{n',k'} \omega_t^{(h|1)} (n+n'|k+k'-2) \nonumber \\
& & \hspace{12mm} \times k' \sum_{m, \ell} \left\{ F^{(h'|1)}(n',m|k',\ell) \omega _{t+1}^{(h'|1)} (m|\ell ) + F^{(h'|1)}(m,n'|\ell ,k') \omega _{t-1}^{(h'|1)} (m|\ell ) \right\} \nonumber \\
& & \left. + \cdots ~\right] + \Delta \zeta _t^{(h|1)} (n|\bar{k}).
\end{eqnarray}
The terms in the first square brackets, similar to that of $\Delta \phi _t(n)$, are due to the stochastic time evolution of a tensor field, relating to the deformation of one square on the surface.
On the other hand, the terms in the second square brackets multiplied by $\lambda_{\rm B}$ are caused by the evolution of the matrix field, or the deformation of one link on the boundary.
The IK-type interaction originated with the latter field extends the boundary loop, according to which the bounded surface is spread, simultaneously creating counterpart disc at a neighboring time. 
Correspondingly to the two types of deformation, we have two constructive noise variables,
\begin{eqnarray}
\label{eq:constructive2}
\Delta \zeta _t^{(h|1)} (\bar{n}|k) & \equiv & {n \over N^{h+1}} \left\{ \left( {A_t \over N} \right)^{n-1}_{dcba} {\rm tr} \left( {\psi_t \over \sqrt{N} } \right) ^k \right\} (\Delta \xi_t )_{abcd}, \nonumber \\
\Delta \zeta _t^{(h|1)} (n|\bar{k}) & \equiv & {k \over N^{h+{1 \over 2}}} \left\{ \left( {A_t \over N} \right)^{n} \left( {\psi_t \over \sqrt{N} } \right) ^{k-1}_{i_2 i_1} \right\}_a \left( (\Delta \eta_t )_a \right)_{i_1 i_2}.
\end{eqnarray}
The difference of the over-lined variable in the l.h.s. expresses for which field a white noise is substituted.
The former relates to the merging interaction of the open surface at a square and the latter does at a link.
Nontrivial correlations concerning the above three constructive noise variables are given by
\begin{eqnarray}
\label{eq:correlationzeta}
\langle \Delta \zeta _t^{(h)} (n)  \Delta \zeta _{t'}^{(h')} (m) \rangle _{\xi} & = & 2\Delta \tau \delta _{tt'} {1 \over N^2} nm \phi _t^{(h+h')} (n+m-2), \nonumber \\
\langle \Delta \zeta _t^{(h|1)} (\bar{n}|k)  \Delta \zeta _{t'}^{(h'|1)} (\bar{m}|\ell ) \rangle _{\xi} & = & 2\Delta \tau \delta _{tt'} {1 \over N^2} nm \omega _t^{(h+h'|2)} (n+m-2|k, \ell ), \nonumber \\
\langle \Delta \zeta _t^{(h)} (n)  \Delta \zeta _{t'}^{(h'|1)} (\bar{m}|\ell) \rangle _{\xi} & = & 2 \Delta \tau \delta _{tt'} {1 \over N^2} nm \omega _t^{(h+h'|1)} (n+m-2|\ell ), \nonumber \\
\langle \Delta \zeta _t^{(h|1)} (n| \bar{k})  \Delta \zeta _{t'}^{(h'|1)} (m|\bar{\ell}) \rangle _{\eta} & = & 2 \lambda_{\rm B} \Delta \tau \delta _{tt'} {1 \over N} k\ell \omega _t^{(h+h'|1)} (n+m|k+\ell -2),
\end{eqnarray}
while the other combination, as well as one noise variable, vanishes.
The first, second and third correlations cause the merging interactions of two closed surfaces, that of two open surfaces and that of open-closed surfaces, respectively, by gluing squares on surfaces.
The last one brings on the merging interaction of two open surfaces at some links of both boundary loops.

We define effective fields as abbreviated forms for the sets of the created fields at the neighboring times multiplied by a one-step propagator:
\begin{eqnarray}
\label{eq:tilde1}
\tilde{\phi}_t^{(h)} (n') & \equiv & \sum_{m=0}^{\infty} \left\{ n' G^{(h)}(n',m) \phi _{t+1}^{(h)} (m) + n' G^{(h)}(m,n') \phi _{t-1}^{(h)} (m) \right\}, \nonumber \\
\tilde{\omega}_t^{(h|1)} (n'|k') & \equiv & \sum_{m=0}^{\infty} \sum_{\ell =0}^{\infty} \left\{ k' F^{(h|1)}(n',m|k',\ell) \omega _{t+1}^{(h|1)} (m|\ell) + k' F^{(h|1)}(m,n'|\ell ,k') \omega _{t-1}^{(h|1)} (m|\ell ) \right\}, \nonumber \\
\end{eqnarray}
to make the concerning terms take the original form of the IK-type interactions.
Later, we assume the terms of the IK-type interaction to remain in the continuum limit.

The stochastic time evolution for the expectation value of any observable $O(\phi, \omega)$ with the noise correlations, is given by $\langle O( \phi _t , \omega _t ; \tau =\tau) \rangle _{\xi \eta} \equiv \langle e^{-\tau H_{\rm FP}} O(\phi _t ,\omega _t ; \tau =0) \rangle _{\xi \eta } $.
In the step of the discrete stochastic time $\Delta \tau$, the minimum evolution $\langle \Delta O(\phi , \omega ) \rangle _{\xi \eta} \equiv  - \Delta \tau \langle H_{\rm FP} O(\phi , \omega ) \rangle _{\xi \eta} $ provides the FP Hamiltonian $H_{\rm FP}$.
With Eqs.(\ref{eq:langevinphi}), (\ref{eq:langevinomega}) and (\ref{eq:correlationzeta}), it is derived through
\begin{eqnarray}
\label{eq:FPH}
-\Delta \tau H_{\rm FP} & =  & \sum_t \sum_{h=0}^{\infty} \sum_{n=1}^{\infty}  \langle \Delta \phi ^{(h)}_t (n) \rangle_{\xi } ~\pi ^{(h)}_t (n) 
 + \sum_t \sum_{h=0}^{\infty} \sum_{n=1}^{\infty} \sum_{k=1}^{\infty} \langle \Delta \omega ^{(h|1)}_t (n|k) \rangle_{\xi \eta} ~\pi ^{(h|1)}_t (n|k)   \nonumber \\
&& + {1 \over 2} \sum_{t,t'} \sum_{h,h'}^{\infty} \sum_{n,m}^{\infty}  \langle \Delta \phi ^{(h)}_t (n) \Delta \phi ^{(h')}_{t'} (m) \rangle_{\xi} ~\pi ^{(h)}_t (n) \pi ^{(h')}_{t'} (m) \nonumber \\
& &  + {1 \over 2} \sum_{t,t'} \sum_{h,h'=0}^{\infty} \sum_{n,m=1}^{\infty}  \sum_{k,\ell=1}^{\infty} \langle \Delta \omega ^{(h|1)}_t (n|k)  \Delta \omega ^{(h'|1)}_{t'} (m|\ell) \rangle_{\xi \eta} ~\pi ^{(h|1)}_t (n|k) \pi ^{(h'|1)}_{t'} (m|\ell) \nonumber \\
& &  + \sum_{t,t'} \sum_{h,h'=0}^{\infty} \sum_{n,m=1}^{\infty}  \sum_{\ell=1}^{\infty} \langle \Delta \phi ^{(h)}_t (n)  \Delta \omega ^{(h'|1)}_{t'} (m|\ell) \rangle_{\xi \eta} ~\pi ^{(h)}_t (n) \pi ^{(h'|1)}_{t'} (m|\ell) \nonumber \\
& & + \cdots ,
\end{eqnarray} 
up to the lowest order of $\Delta \tau$.
The FP Hamiltonian is a functional of the fields $\phi_t^{(h)}(n)$, $\omega_t^{(h|1)}(n|k)$, $\omega_t^{(h|2)}(n|k,k')$ and the derivatives with respect to them.
The dots in the last line contain terms deforming surfaces with higher boundary number, whose leading terms we need later in the estimation of the commutation relations. 
We have denoted the annihilation operators corresponding to these fields,
\begin{eqnarray}
\label{eq:annihilation}
\pi_t^{(h)} (n) & \equiv & {\partial \over \partial \phi_t^{(h)} (n)}, \nonumber \\
\pi_t^{(h|1)} (n|k) & \equiv & {\partial \over \partial \omega_t^{(h|1)} (n|k)}, \nonumber \\
\pi_t^{(h|2)} (n|k,k') & \equiv & {\partial \over \partial \omega _t^{(h|2)} (n|k,k')},
\end{eqnarray}
hence they satisfy the following commutation relations:
\begin{eqnarray}
\label{eq:commutator}
\left[ \pi_t^{(h)} (n) , \phi_{t'}^{(h')} (n') \right]  & = & \delta _{tt'} \delta _{hh'} \delta_{nn'}, \nonumber \\
\left[ \pi_t^{(h|1)} (n|k) , \omega_{t'}^{(h'|1)} (n'|k') \right]  & = & \delta _{tt'} \delta _{hh'} \delta_{nn'} \delta_{kk'}, \nonumber \\ 
\left[ \pi_t^{(h|2)} (n|k,\ell) , \omega_{t'}^{(h'|2)} (n'|k',\ell') \right]  & = & {1 \over 2} \delta _{tt'} \delta _{hh'} \delta_{nn'} ( \delta_{kk'} \delta_{\ell \ell'} +\delta_{k \ell'} \delta_{\ell k'}).
\end{eqnarray}
We will describe the explicit form of the continuum FP Hamiltonian for the estimation of the scaling order in the next section and further that of the discrete generators for the investigation of the commutation relation in section \ref{sec:algebra}.
Hence, we refrain here from writing them down to avoid the repetition. 
Now, we are sure that for the surface field with arbitrary numbers of handles and loops, $\omega_t^{(h|b)}(n|k_1,k_2,\cdots,k_b)$, the deformation of the fields possesses common structure  from two origins, at one square and at one link.
The FP Hamiltonian contains infinite series of generators multiplied by an annihilation operator, $\pi_t^{(h|b)}(n|k_1,k_2,\cdots,k_b)$.
However, for the higher numbers of handles $h$ and boundary loops $b$ of a surface, we expect the concerning interaction to scale out in the continuum limit.

\section{Continuum limit}\label{sec:continuum}

We will take the double-scaling limit to obtain CDT surface field theory from the above discrete model including some uninvited CDT breaking interactions.
We assume all kinds of IK-type interaction to remain with the scaling of $a^0$, zero-th power of the scaling parameter, as the possible quantum effects in CDT.
The unit length of both time-like link and space-like link, or the minimal length and time, is $a$ according to CDT.
At the double scaling limit, $a \rightarrow 0$, the square number $n$, the boundary link number $k$ and the time step $t$ of CDT scale to the area of a surface $A$, the length of a boundary loop $L$ and time $T$, respectively, as
\begin{eqnarray}
\label{eq:continuum}
A \equiv a^2 n, \hspace{10mm} L \equiv ak, \hspace{10mm} T \equiv at. 
\end{eqnarray}
The coupling constants $g$ and $g_{\rm B}$ relate to the cosmological constant $\Lambda$ and the boundary cosmological constant $\Lambda _{\rm B}$, respectively, by
\begin{eqnarray}
\label{eq:cosmological}
g \equiv {1 \over 2} e^{-\Lambda {a^3}}, \hspace{10mm}  g_{\rm B} \equiv {1 \over 2} e^{-\Lambda_{\rm B} {a^2}} .
\end{eqnarray}
Now, we introduce two scaling dimensions $D$ and $D_N$ to estimate the scaling behavior of each term in the FP Hamiltonian.
The latter is defined in the scaling of $1/N^2$ to a constant $G_{\rm st}$ as
\begin{eqnarray}
\label{eq:coupling}
G_{\rm st} \equiv a^{D_N}{1 \over N^2} ,
\end{eqnarray}
where the subscript character ``st'' refers to the string coupling constant, which is the counterpart in the 2D model.
On the other hand, the former defines the scaling of the closed surface field, whose creation and annihilation operators are assumed to scale as
\begin{eqnarray}
\label{eq:sphere}
\Phi (A;T) \equiv a^{-{1 \over 2}D} \phi _t (n) ,\hspace{10mm} \Pi (A;T) \equiv a^{{1 \over 2}D-3} \pi _t (n) .
\end{eqnarray}
For closed surfaces with one and $h$ handles we define, respectively,
\begin{eqnarray}
\label{eq:closed surface field}
\Phi ^{(1)} (A;T) \equiv a^{-{1 \over 2}D+{1 \over 2}D_N} \phi _t^{(1)} (n), \hspace{10mm}  \Pi ^{(1)} (A;T) \equiv a^{{1 \over 2}D-{1 \over 2}D_N -3} \pi _t^{(1)} (n), \nonumber \\
\Phi ^{(h)} (A;T) \equiv a^{-{1 \over 2}D+{h \over 2}D_N} \phi _t^{(h)} (n), \hspace{10mm}  \Pi ^{(h)} (A;T) \equiv a^{{1 \over 2}D-{h \over 2}D_N -3} \pi _t^{(h)} (n).
\end{eqnarray}
We have counted the multiplication of $a^{{1 \over 2}D_N}$ for the factor $1/N$ corresponding to each handle, in the discrete fields (\ref{eq:closed}) (and (\ref{eq:open2}) for the open surface, as well).
Above scaling indices of annihilation operators are fixed by the commutation relation of the continuum form,
\begin{eqnarray}
\label{eq:commutator1}
\left[ \Pi ^{(h)} (A;T), \Phi ^{(h')} (A';T') \right] & = & \delta_{h,h'} \delta (A-A') \delta (T-T'),
\end{eqnarray}
with the first line of Eq.(\ref{eq:commutator}).
In order to make the IK-type interaction of the closed surfaces, the third line of Eq.(\ref{eq:langevinphi}), remain in the scaling, we have to define the continuum stochastic time as
\begin{eqnarray}
\label{eq:stochastic-time}
d \tau & \equiv & a^{{1 \over 2}D-4} \Delta \tau .
\end{eqnarray}
For the existence of the IK-type interaction of the closed surface with open surfaces, the fourth and fifth lines of Eq.(\ref{eq:langevinphi}), we define the continuum version of creation and annihilation operators of the simplest open surface, the disc (an open surface with one boundary loop), as
\begin{eqnarray}
\label{eq:disc1}
\Omega (A|L;T) \equiv a^{-{1 \over 2}D-{1 \over 4}D_N-1} \omega _t (n|k), \hspace{10mm}  \Pi^{(0|1)} (A|L;T) \equiv a^{{1 \over 2}D+{1 \over 4}D_N-3} \pi^{(0|1)} _t (n|k). 
\end{eqnarray}
The scaling of the latter operator is decided from the commutation relation,
\begin{eqnarray}
\label{eq:commutator2}
\left[ \Pi ^{(h|1)} (A|L;T), \Omega ^{(h'|1)} (A'|L';T') \right] & = & \delta_{h,h'} \delta (A-A') \delta (L-L') \delta (T-T'),
\end{eqnarray}
and the second equation of (\ref{eq:commutator}).
With the scaling dimension fixing until now, the IK-type interaction of an open surface related to a closed surface, the third and fourth lines of Eq.(\ref{eq:langevinomega}), is guaranteed to exist as well.
In the similar way, the field operators of an open surface with two boundary loops, or a cylinder field, are decided with the condition that the IK-type interaction of a disc changing to a cylinder, the fifth and sixth lines of Eq.(\ref{eq:langevinomega}), remains,
\begin{eqnarray}
\label{eq:cylinder1}
\Omega ^{(0|2)} (A|L,L';T) \equiv a^{-{1 \over 2}D-{1 \over 2}D_N-2} \omega _t^{(0|2)} (n|k,k'), && \Pi ^{(0|2)} (A|L,L';T) \equiv a^{{1 \over 2}D+{1 \over 2}D_N -3} \pi _t^{(0|2)} (n|k,k'). \nonumber \\
\end{eqnarray}
Here, we have utilized the commutation relation
\begin{eqnarray}
\label{eq:commutator3}
&&\left[ \Pi^{(h|2)}  (A|L_1,L_2;T), \Omega ^{(h'|2)} (A'|L'_1,L'_2;T') \right] \nonumber \\
&& \hspace{5mm} = {1 \over 2} \delta_{h,h'} \delta (A-A') \left\{ \delta (L_1-L'_1) \delta (L_2-L'_2) + \delta (L_1-L'_2) \delta (L_2-L'_1) \right\} \delta (T-T') ,
\end{eqnarray}
corresponding to the last equation of (\ref{eq:commutator}).
Since one loop addition seems to relate to the factor $a^{-{1 \over 4}D_N-1}$ in the creation operators, the open surface operators with $b$ boundary loops are
\begin{eqnarray}
\label{eq:open}
\Omega ^{(0|b)} (A|L,L',\cdots ;T) \equiv a^{-{1 \over 2}D-{b \over 4}D_N-b} \omega _t^{(0|b)} (n|k,k',\cdots ) , \nonumber \\
\Pi ^{(0|b)} (A|L,L', \cdots ;T) \equiv a^{{1 \over 2}D+{b \over 4}D_N -3} \pi _t^{(0|b)} (n|k,k',\cdots ).
\end{eqnarray}
For a surface with $b$ boundary loops, the commutation relation is
\begin{eqnarray}
\label{eq:commutator4}
&&\left[ \Pi^{(h|b)}  (A|L_1,L_2,\cdots ,L_b;T), \Omega ^{(h'|b')} (A'|L'_1,L'_2,\cdots ,L'_{b'};T') \right] \nonumber \\
&& \hspace{10mm} = {1 \over b!} \delta_{h,h'} \delta_{b,b'} \delta (A-A') \left\{ \delta (L_1-\underline{L'_1}) \delta (L_2-\underline{L'_2}) \cdots \delta (L_b-\underline{L'_b}) \right\} \delta (T-T') ,
\end{eqnarray}
where, in the r.h.s., we sum up all the possible permutation of the length variables with underline.
The effective fields in the IK-type interactions, (\ref{eq:tilde1}), are assumed to take the same scaling as the original fields,
\begin{eqnarray}
\label{eq:tilde2}
\tilde{\Phi} (A;T) \equiv a^{-{1 \over 2}D} \tilde{\phi} _t (n) &,&
\tilde{\Omega} (A|L;T) \equiv a^{-{1 \over 2}D-{1 \over 4}D_N-1} \tilde{\omega} _t (n|k).
\end{eqnarray}
It is realized with the scaling of one-step propagators as
\begin{eqnarray}
\label{1step}
\tilde{G}^{(h)} (A , A') & \equiv & a^{-4} G^{(h)}(n,n'), \nonumber \\
\tilde{F}^{(h|1)} (A , A'|L,L') & \equiv & a^{-4} F^{(h|1)}(n,n'|k,k').
\end{eqnarray}
At last, to turn on the IK-type interaction for the boundary loop, the tenth and eleventh lines of Eq.(\ref{eq:langevinomega}), we fix the scaling of the boundary parameter of the stochastic evolution,
\begin{eqnarray}
\label{eq:boundary}
\lambda_{\cal B} & \equiv & a^{{1 \over 4}D_N +1} \lambda_{\rm B}.
\end{eqnarray}
Then, the dimensionless relation $H_{\rm FP} \Delta \tau = {\cal H}_{\rm FP} d \tau $ leads to the continuum FP Hamiltonian:
\begin{eqnarray}
{\cal H}_{\rm FP} d \tau &\hspace{-3mm}=& - d\tau \int dT \int_0^{\infty} dA A \left[- a^{-{1 \over 2}D+5}  \Lambda \Phi (A;T) \right. \label{eq:FPH1} \\
&& + a^{-{1 \over 2}D-{D_N}} \sqrt{G_{\rm st}} A \Phi ^{(1)} (A; T)  \label{eq:FPH2} \\
&& + a^{-{1 \over 2}D-{1 \over 4}D_N+3} \sqrt{G_{\rm st}} \left( {1 \over 2} \Omega(A|2a;T) + {1 \over 4} \Omega(A|4a;T) \right) \label{eq:FPH3} \\
&& + \int_0^{\infty} dA' \Phi (A+A';T) \tilde{\Phi} (A';T) \label{eq:FPH4} \\
&& + \sqrt{G_{\rm st}} \int_0^{\infty} dA' \int_0^{\infty} dL'{A' \over L'} \Omega(A+A'|L';T) \tilde{\Omega}(A'|L';T) \label{eq:FPH5} \\
&& + a^{-D-D_N+1} G_{\rm st} \int_0^{\infty} dA' A'  \Phi  (A+A';T) \Pi  (A';T) \label{eq:FPH6} \\
&& \left. + a^{-D-D_N+1} G_{\rm st} \int_0^{\infty} dA' \int_0^{\infty} dL' A' \Omega (A+A'|L';T) \Pi^{(0|1)} (A'|L';T) \right] \Pi (A;T) \label{eq:FPH7} \\
\nonumber \\
&& -  d \tau \int dT \int_0^{\infty} dA \int_0^{\infty} dL  \left[ A \left\{ - a^{-{1 \over 2}D+5} \Lambda \Omega(A|L;T) \right. \right. \label{eq:FPH8} \\
&& + a^{-{1 \over 2}D-D_N} \sqrt{G_{\rm st}} A \Omega^{(1|1)}(A|L;T) \label{eq:FPH9} \\
&& + a^{-{1 \over 2}D-{1 \over 4}D_N +3} \sqrt{G_{\rm st}} \left( {1 \over 2} \Omega^{(0|2)}(A|L,2a;T) + {1 \over 4} \Omega^{(0|2)}(A|L,4a;T) \right) \label{eq:FPH10} \\
&& + \int_0^{\infty} dA' \Omega (A+A'|L;T) \tilde{\Phi} (A';T) \label{eq:FPH11} \\
&& + \sqrt{G_{\rm st}} \int_0^{\infty} dA' \int_0^{\infty} dL' {A' \over L'} \Omega^{(0|2)} (A+A'|L,L';T) \tilde{\Omega} (A'|L';T) \label{eq:FPH12} \\
&& + a^{-D-D_N+1} G_{\rm st} \int_0^{\infty} dA' A' \Omega (A+A'|L;T) \Pi (A';T) \label{eq:FPH13} \\
&& \left. + a^{-D-D_N +1} G_{\rm st}  \int_0^{\infty}  dA'  \int_0^{\infty}  dL' A' \Omega^{(0|2)}  (A+A'|L,L';T) \Pi ^{(0|1)}  (A'|L';T) \right\} \label{eq:FPH14} \\
\nonumber \\
&& + \lambda_{\cal B} L \left\{ a^{-{1 \over 2}D-{1 \over 4}D_N+3} ~2 {\partial \over \partial L} \Omega (A|L;T) \label{eq:FPH15} \right.  \\
&& + a^{-{1 \over 2}D-{1 \over 2}D_N +2} \sqrt{G_{\rm st}} \int_0^L dL' \Omega ^{(0|2)} (A|L,L';T) \label{eq:FPH16} \\
&& + \int_0^{\infty} dA' \int_0^{\infty} dL' \Omega (A+A'|L+L';T) \tilde{\Omega}(A'|L';T)  \label{eq:FPH17} \\
&& \left. \left. \!\! + a^{-D-D_N +1} \sqrt{G_{\rm st}} \int_0^{\infty} \!\!\!\! dA' \!\! \int_0^{\infty} \!\!\!\! dL' L' \Omega (A + A'|L + L'; T) \Pi^{(0|1)} (A'|L';T) \right\} \right] \Pi^{(0|1)} (A|L;T) . \nonumber \\
\label{eq:FPH18}~
\end{eqnarray}
\begin{figure}
$\bullet $ Deformation of closed surface
\vspace{-6mm}
\\
\vspace{-1mm}
 \begin{minipage}{0.24\hsize}
  \begin{center}
   \scalebox{1}{\includegraphics[width=40mm, bb=0 0 960 720]{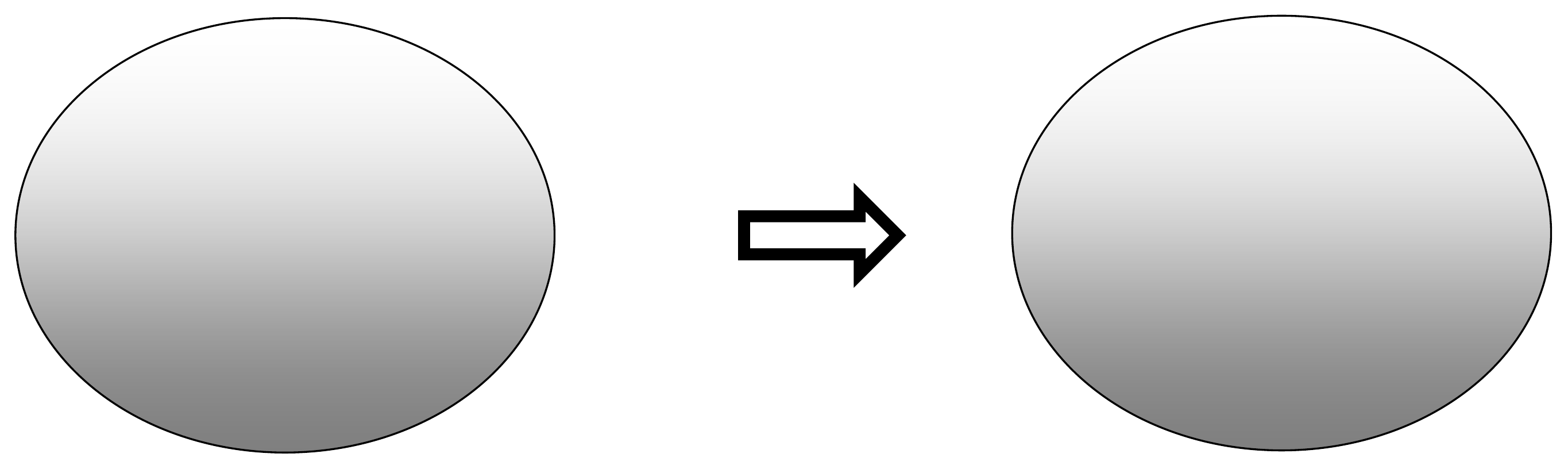}}
(\ref{eq:FPH1})~~Propagation\\
 in a time: $\times$
\end{center}
 \end{minipage}
 \begin{minipage}{0.24\hsize}
  \begin{center}
   \scalebox{1}{\includegraphics[width=40mm, bb=0 0 960 720]{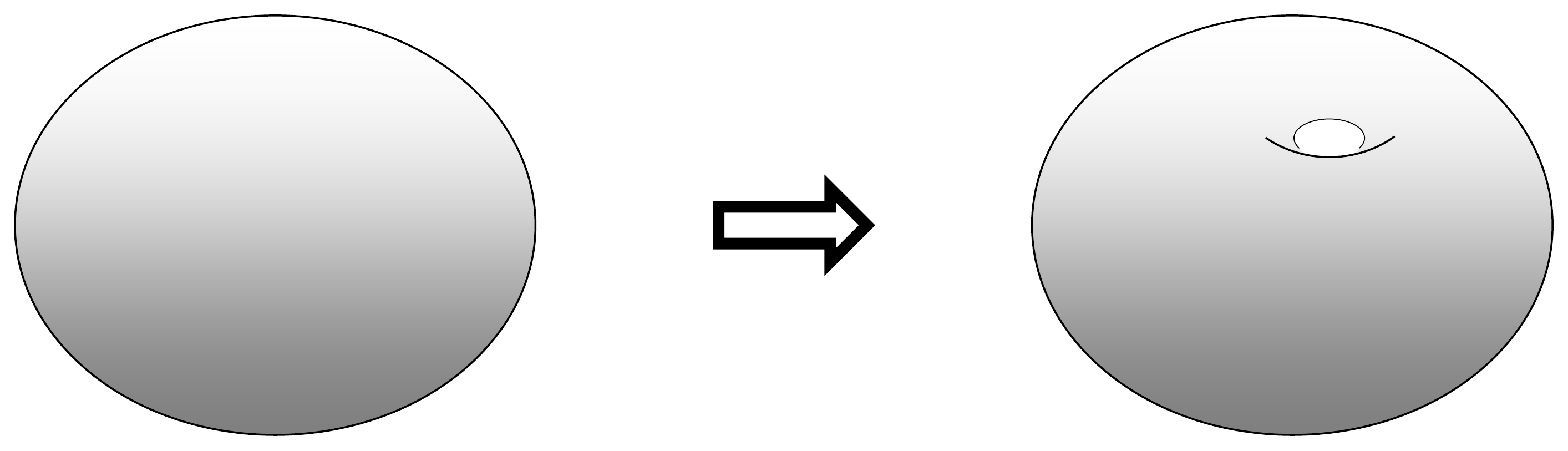}}
(\ref{eq:FPH2})~~Handle-adding: \\
 $\times$
\end{center}
 \end{minipage}
 \begin{minipage}{0.24\hsize}
  \begin{center}
   \scalebox{1}{\includegraphics[width=40mm, bb=0 0 960 720]{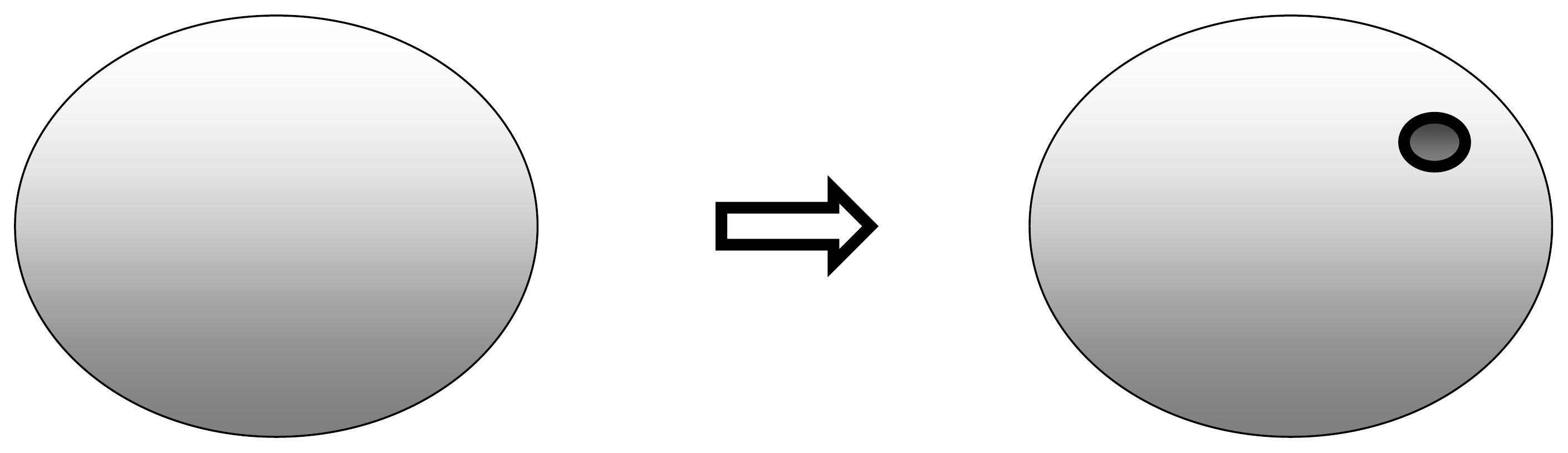}}
(\ref{eq:FPH3})~~Baby loop-\\
adding: $\times$
\end{center}
 \end{minipage}
 \begin{minipage}{0.24\hsize}
  \begin{center}
   \scalebox{1}{\includegraphics[width=40mm, bb=0 0 960 720]{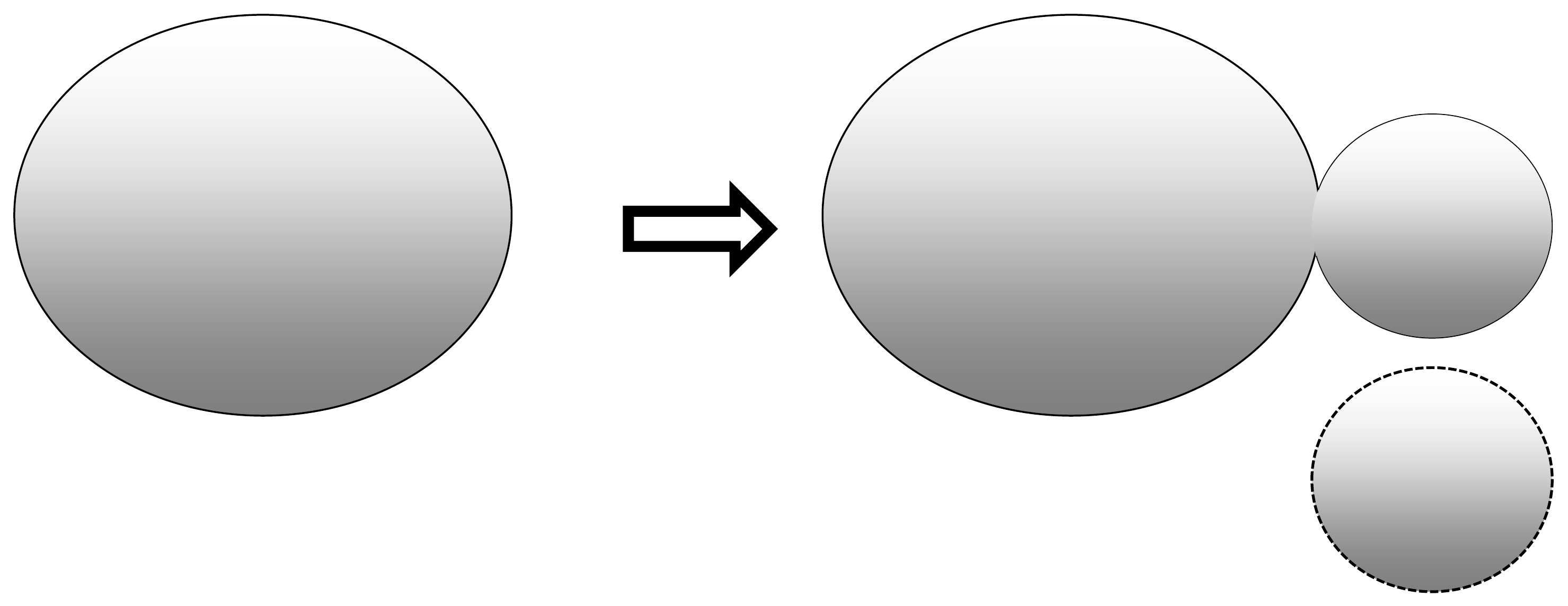}}
(\ref{eq:FPH4})~~IK-type with\\
 closed surface: $\bigcirc$
\end{center}
 \end{minipage}
\vspace{-6mm}
\\
\vspace{-4mm} 
\begin{minipage}{0.24\hsize}
  \begin{center}
   \scalebox{1}{\includegraphics[width=40mm, bb=0 0 960 720]{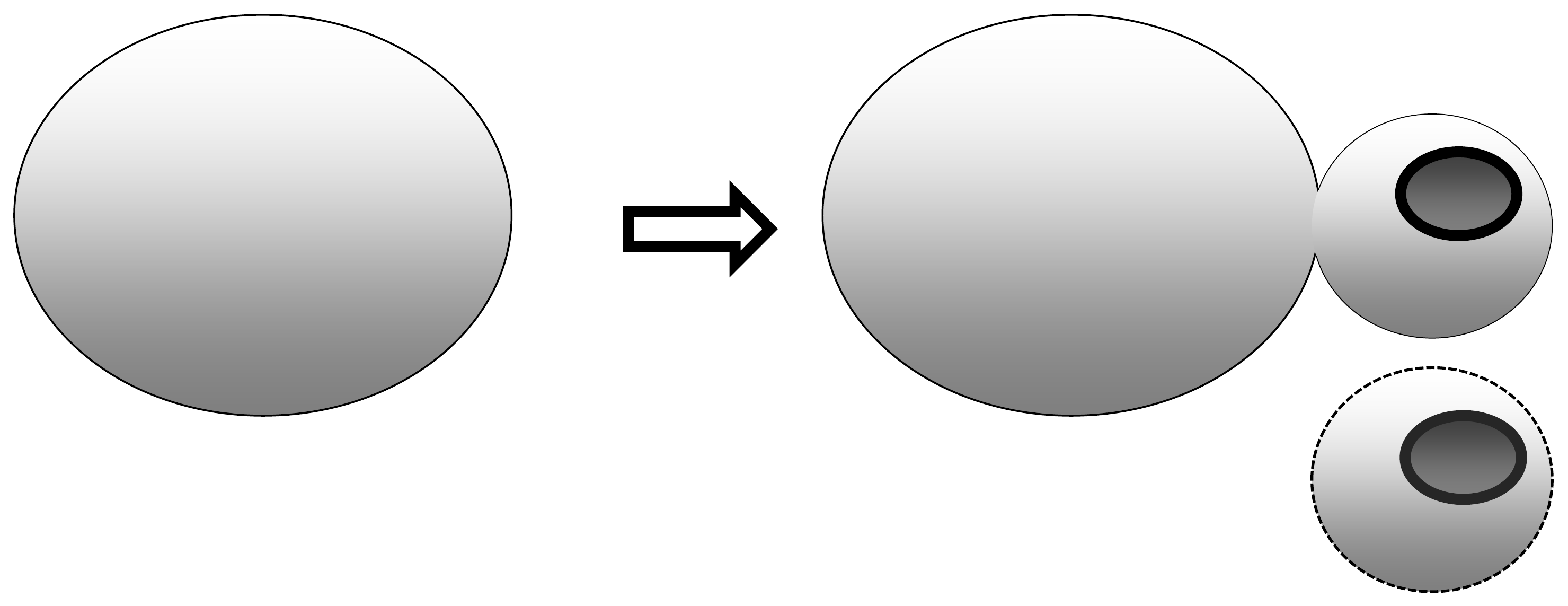}}
(\ref{eq:FPH5})~~IK-type with\\
 open surface: $\bigcirc$
\end{center}
 \end{minipage}
 \begin{minipage}{0.24\hsize}
  \begin{center}
   \scalebox{1}{\includegraphics[width=40mm, bb=0 0 960 720]{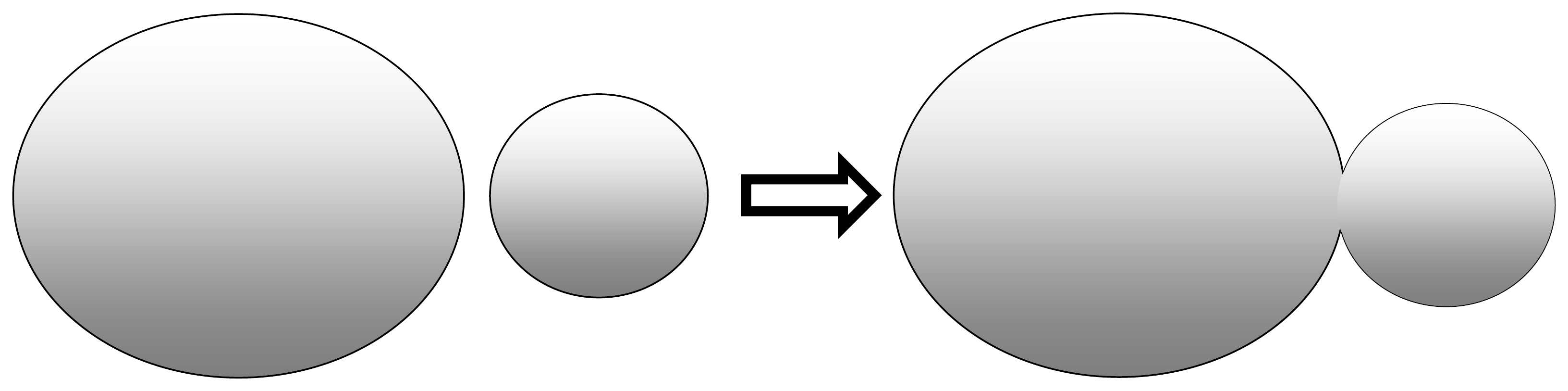}}
(\ref{eq:FPH6})~~Merging with\\
 closed surface: $\times$
\end{center}
 \end{minipage}
 \begin{minipage}{0.24\hsize}
  \begin{center}
   \scalebox{1}{\includegraphics[width=40mm, bb=0 0 960 720]{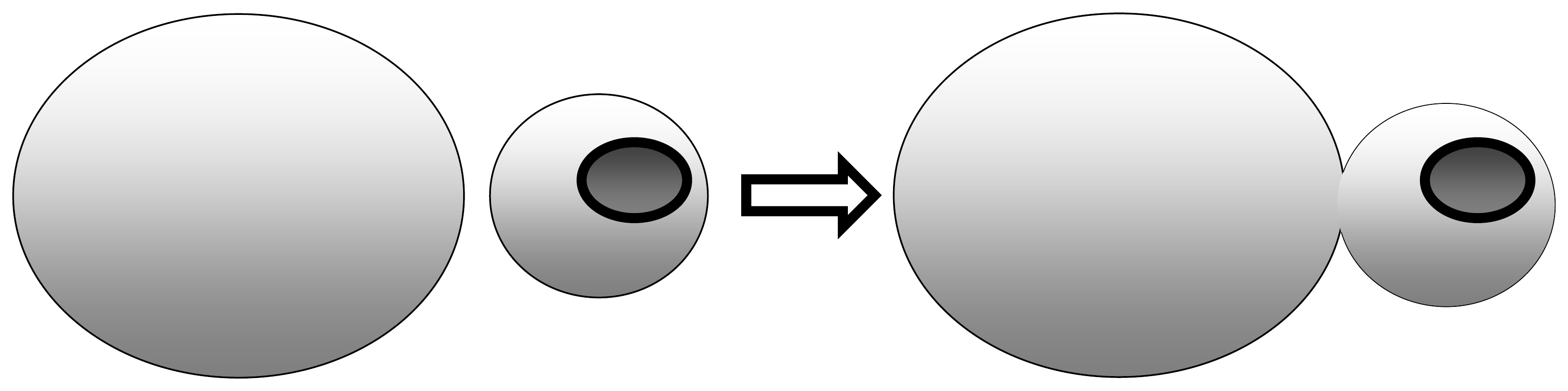}}
(\ref{eq:FPH7})~~Merging with\\
 open surface: $\times$
\end{center}
 \end{minipage}
\vspace{16mm}
\\
$\bullet $ Deformation on open surface
\vspace{-6mm}
\\
\vspace{-10mm}
 \begin{minipage}{0.24\hsize}
  \begin{center}
   \scalebox{1}{\includegraphics[width=40mm, bb=0 0 960 720]{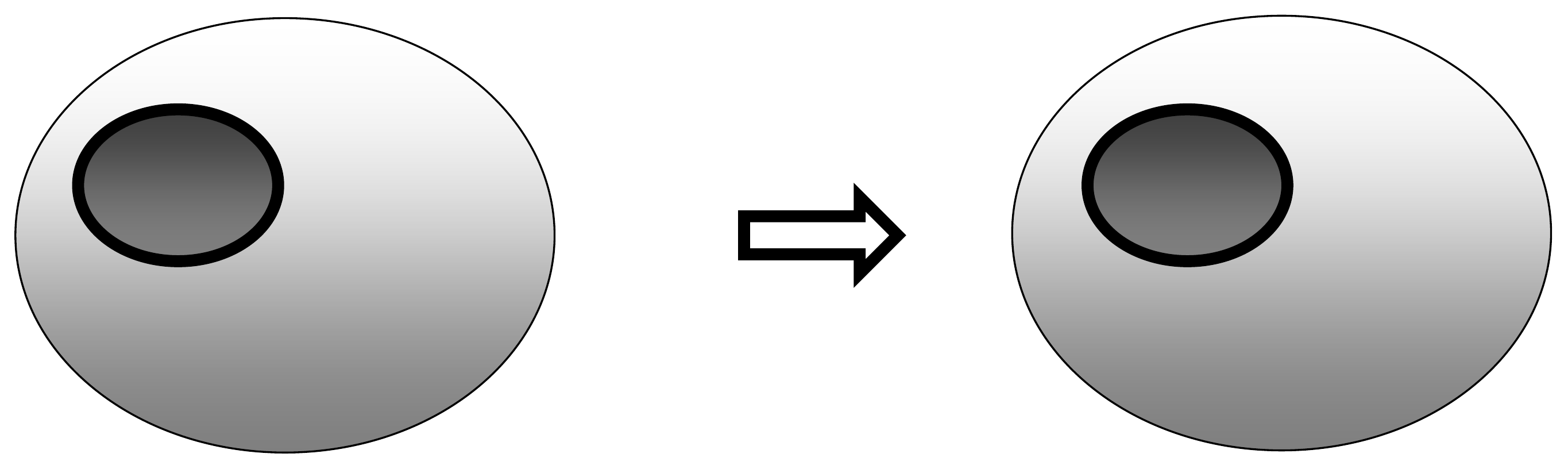}}
(\ref{eq:FPH8})~~Propagation\\
 in a time: $\times$
\end{center}
 \end{minipage}
 \begin{minipage}{0.24\hsize}
  \begin{center}
   \scalebox{1}{\includegraphics[width=40mm, bb=0 0 960 720]{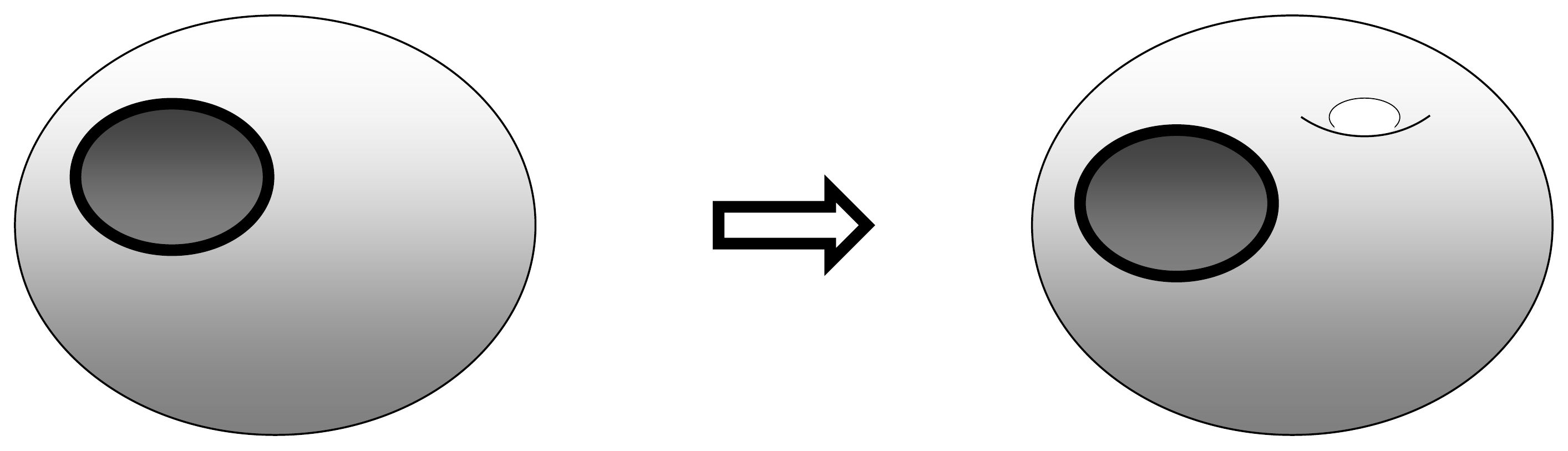}}
(\ref{eq:FPH9})~~Handle-adding:\\
 $\times$
\end{center}
 \end{minipage}
 \begin{minipage}{0.24\hsize}
  \begin{center}
   \scalebox{1}{\includegraphics[width=40mm, bb=0 0 960 720]{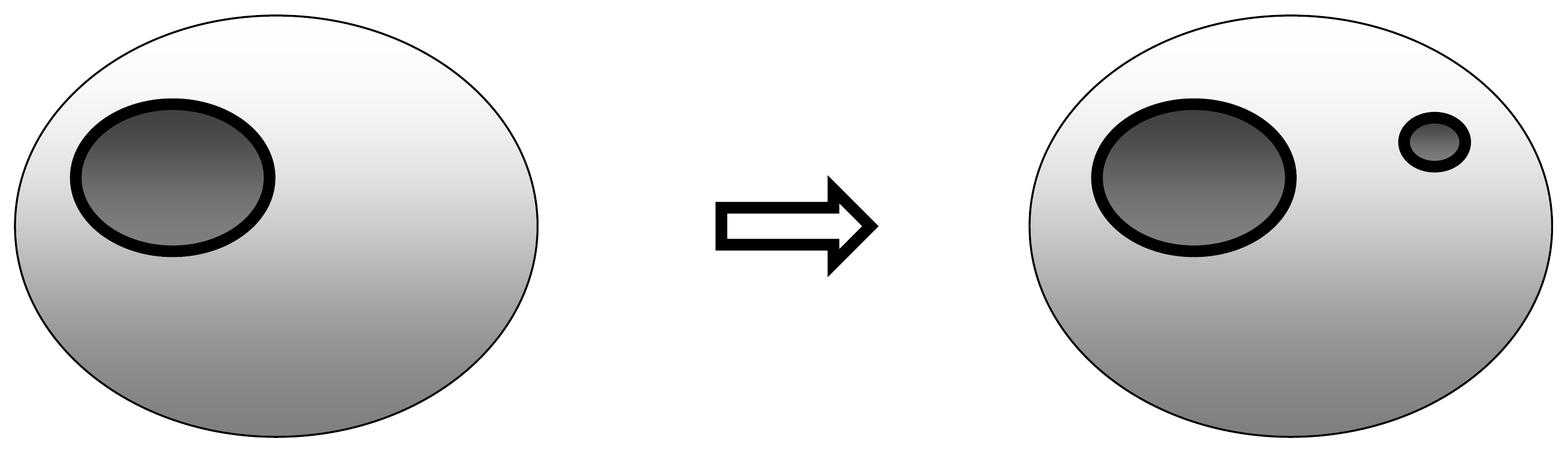}}
(\ref{eq:FPH10})~~Baby loop-\\
 adding: $\times$
\end{center}
 \end{minipage}
 \begin{minipage}{0.24\hsize}
  \begin{center}
   \scalebox{1}{\includegraphics[width=40mm, bb=0 0 960 720]{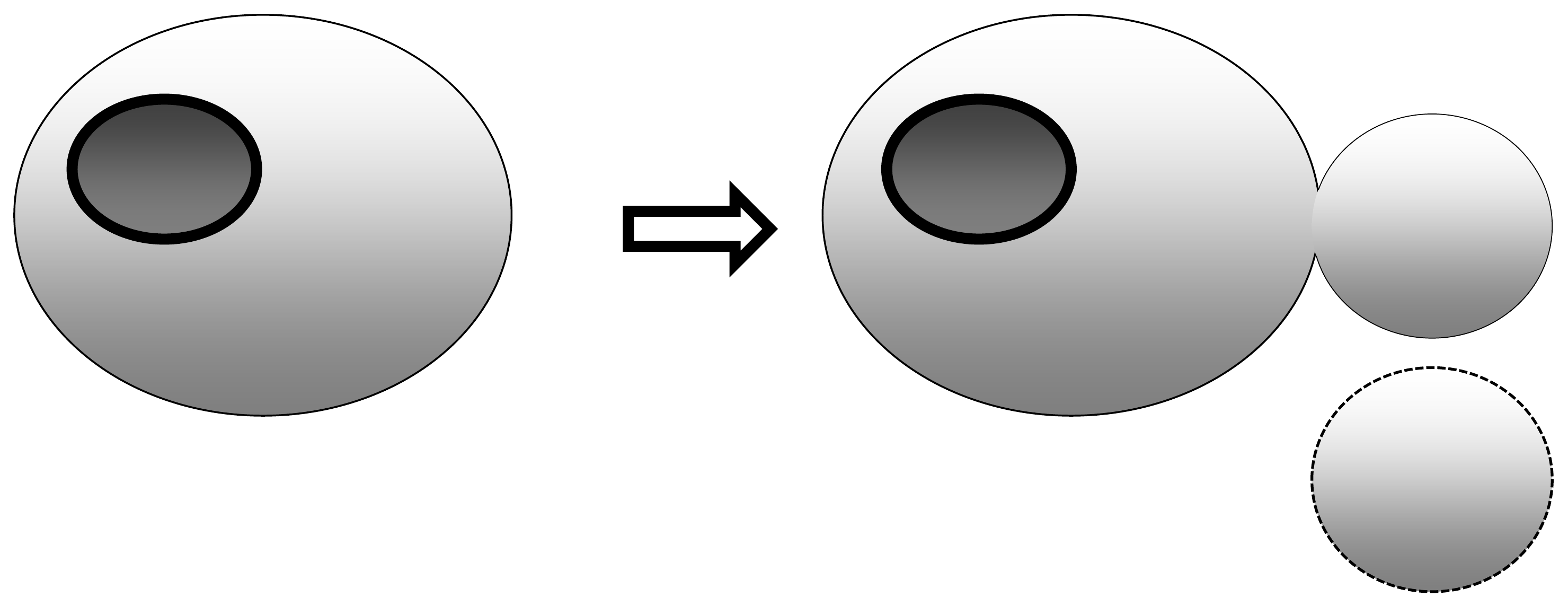}}
(\ref{eq:FPH11})~~IK-type with\\
 closed surface: $\bigcirc$
\end{center}
 \end{minipage}
\vspace{3mm}
\\
 \begin{minipage}{0.24\hsize}
  \begin{center}
   \scalebox{1}{\includegraphics[width=40mm, bb=0 0 960 720]{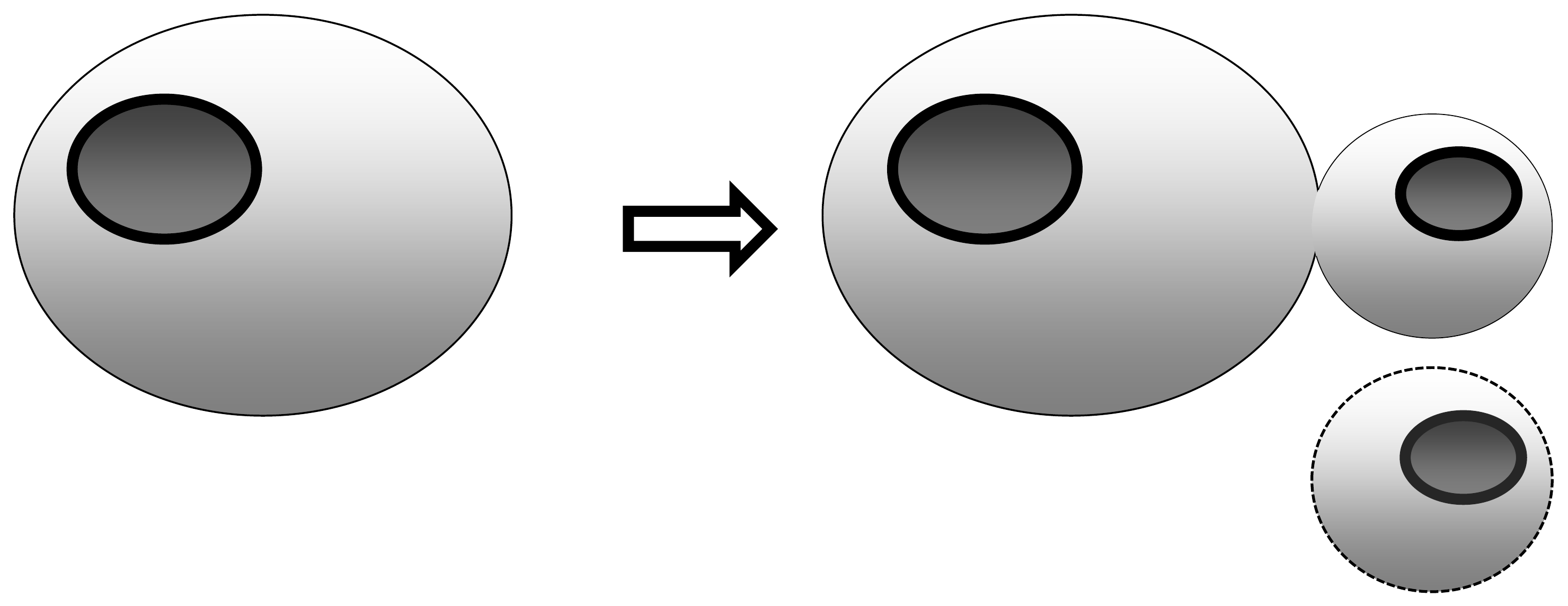}}
(\ref{eq:FPH12})~~IK-type with\\
 open surface: $\bigcirc$
\end{center}
 \end{minipage}
 \begin{minipage}{0.24\hsize}
  \begin{center}
   \scalebox{1}{\includegraphics[width=40mm, bb=0 0 960 720]{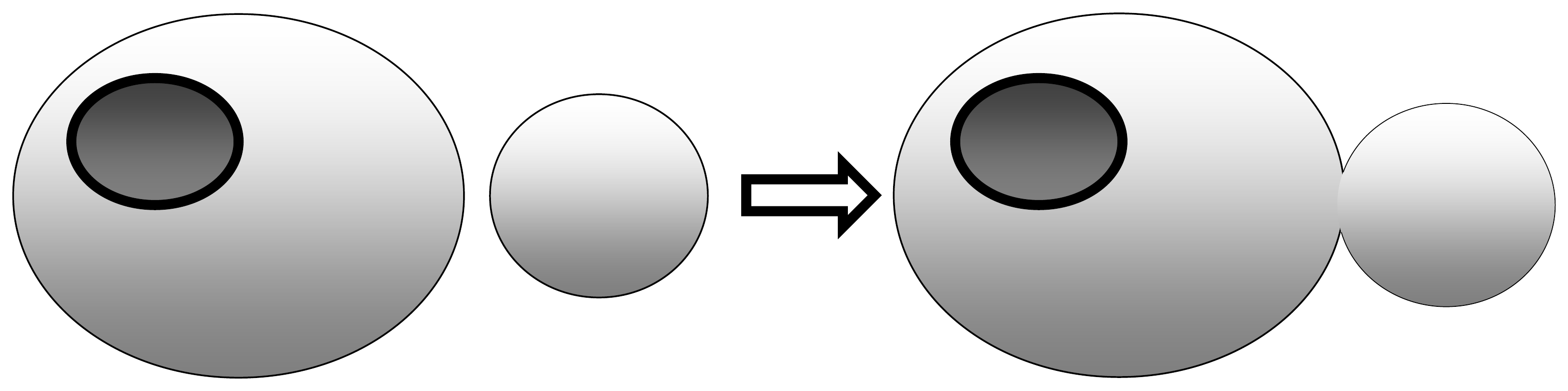}}
(\ref{eq:FPH13})~~Merging with \\
 closed surface: $\times$\\
\end{center}
 \end{minipage}
 \begin{minipage}{0.24\hsize}
  \begin{center}
   \scalebox{1}{\includegraphics[width=40mm, bb=0 0 960 720]{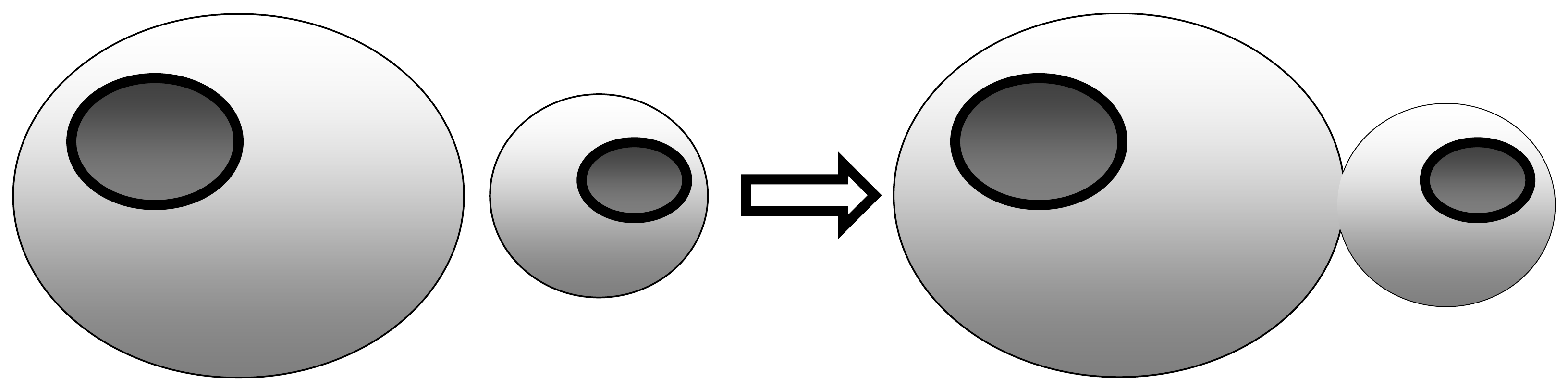}}
(\ref{eq:FPH14})~~Merging with\\
 open surface: $\times$
\end{center}
 \end{minipage}
\vspace{12mm}
\\
$\bullet $ Deformation on the boundary loop of open surface
\vspace{-6mm}
\\
\vspace{-4mm}
 \begin{minipage}{0.24\hsize}
  \begin{center}
   \scalebox{1}{\includegraphics[width=40mm, bb=0 0 960 720]{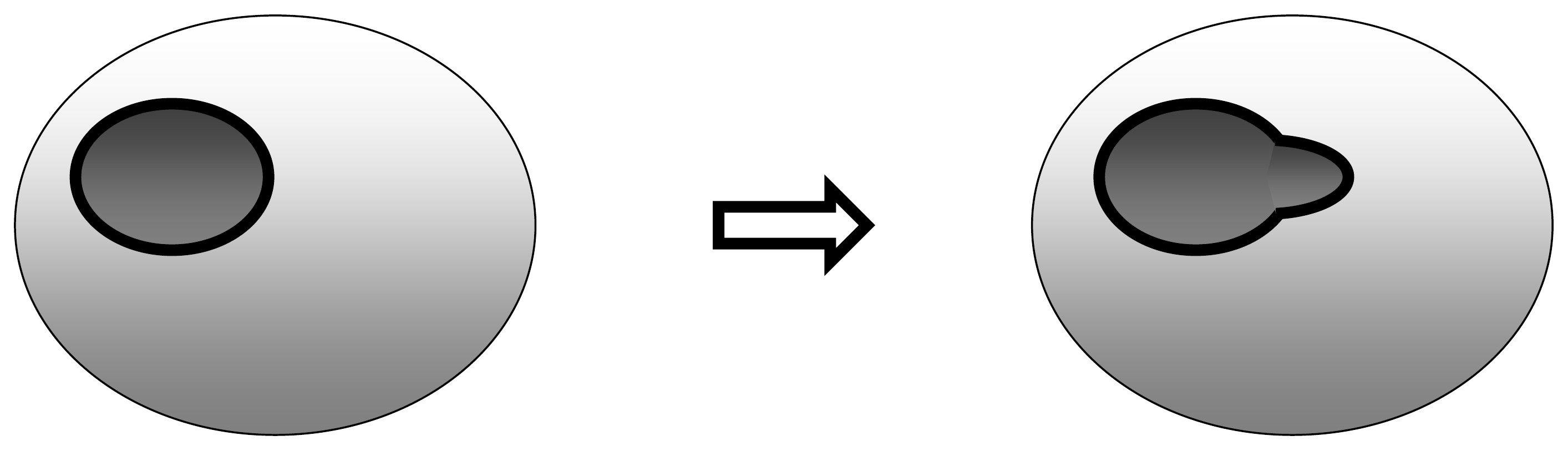}}
(\ref{eq:FPH15})~~Propagation\\
 in a time: $\times$
\end{center}
 \end{minipage}
 \begin{minipage}{0.24\hsize}
  \begin{center}
   \scalebox{1}{\includegraphics[width=40mm, bb=0 0 960 720]{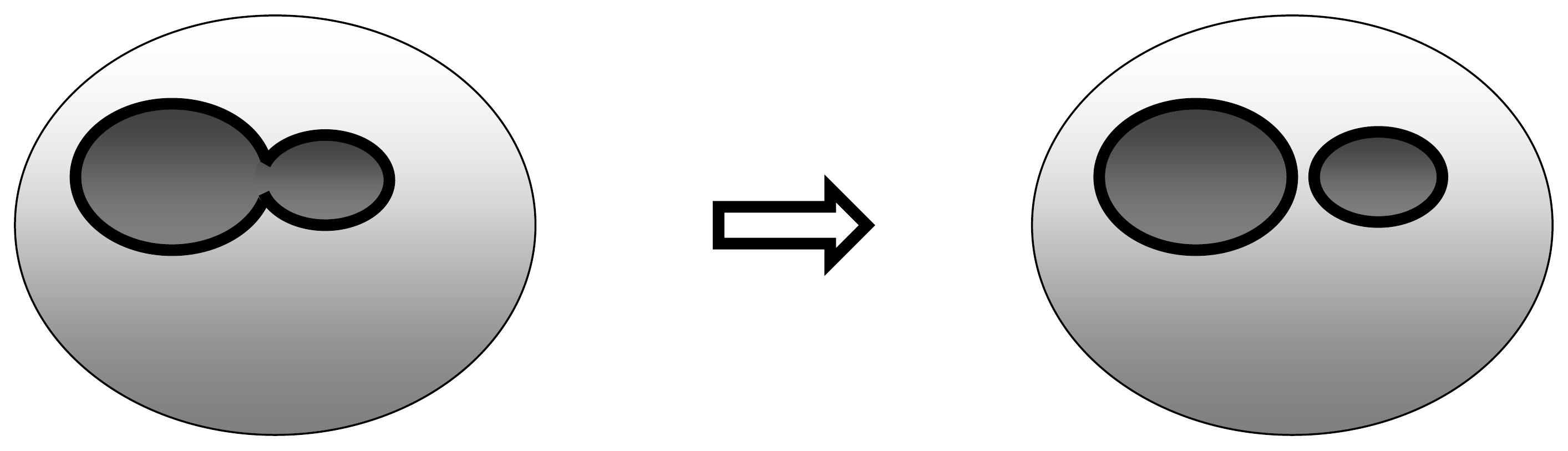}}
(\ref{eq:FPH16})~~Loop splitting: \\
 $\times$
\end{center}
 \end{minipage}
 \begin{minipage}{0.24\hsize}
  \begin{center}
   \scalebox{1}{\includegraphics[width=40mm, bb=0 0 960 720]{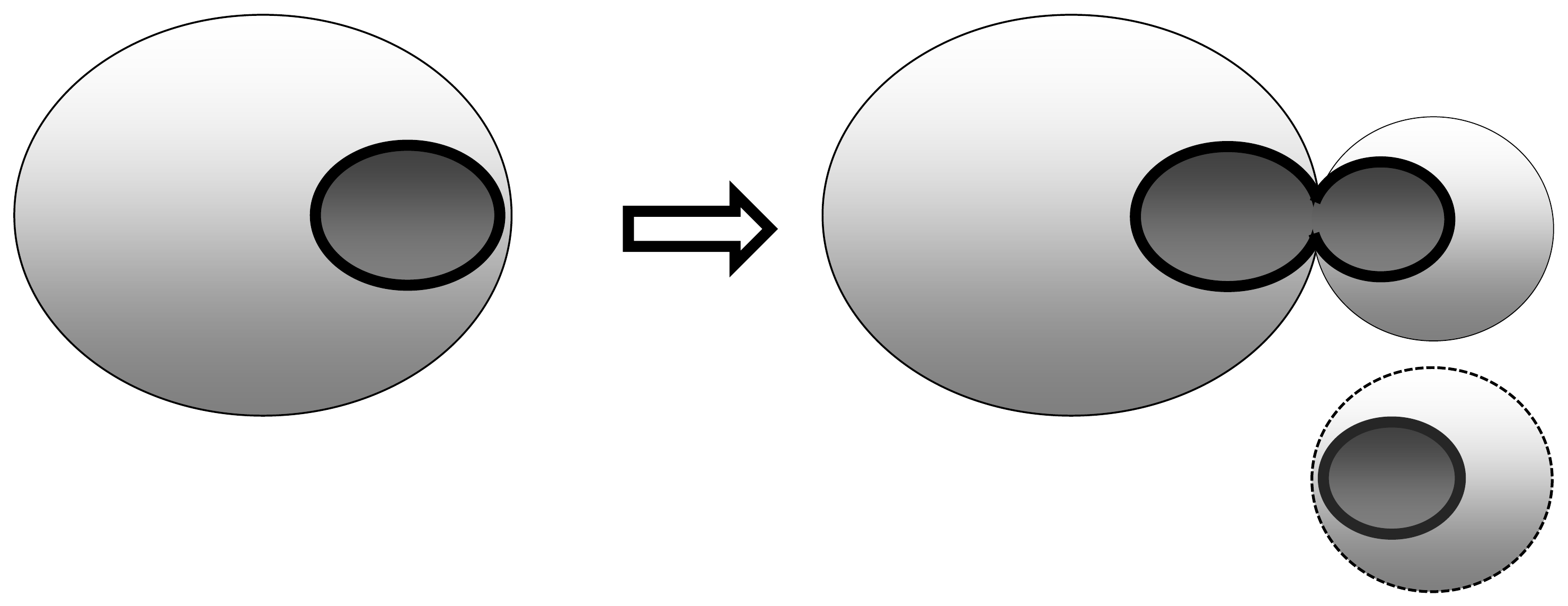}}
(\ref{eq:FPH17})~~IK-type with\\
 loop: $\bigcirc$
\end{center}
 \end{minipage}
 \begin{minipage}{0.24\hsize}
  \begin{center}
   \scalebox{1}{\includegraphics[width=40mm, bb=0 0 960 720]{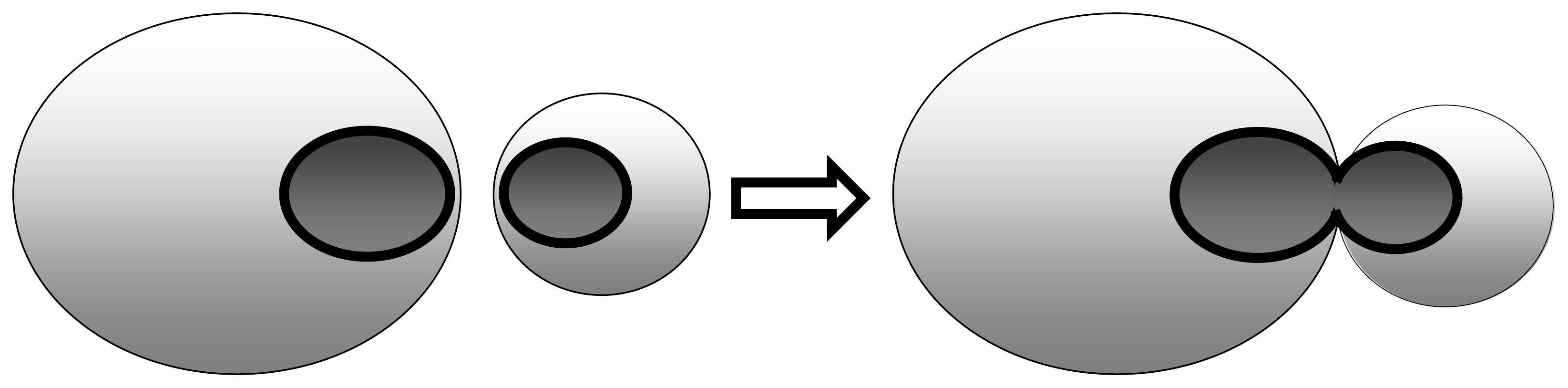}}
(\ref{eq:FPH18})~~Merging with\\
 loop: $\times$
\end{center}
 \end{minipage}
\vspace{8mm}
\caption{
18 processes of the stochastic time evolution of open-closed surface.
Interactions attached with ``$\bigcirc$" survive in ``the closed CDT scaling'', while those with ``$\times$" scale out.
Only the IK-type interactions remain as the quantum effect in 3D CDT surface field theory.
}
\label{fig:Hamiltonian}
\end{figure}
In the above expression, we have not contained the IK-type and merging interactions concerning higher order surfaces, or the surfaces with one or more handles and open surfaces with two or more boundary loops.
These IK-type interactions are suppressed by the higher power of $a$ in the scaling limit, while the merging interactions scale by the same order $a^{-D-D_N+1}$ independent of handle number $h(>0)$ of the object surface
\footnote{
The terms of the IK-type interaction to the surfaces with $h(\ge 1)$ handles, or the next order terms of (\ref{eq:FPH4}),
\begin{eqnarray}
\sum_{h} a^{-hD_N} \int_0^{\infty} dA'  \Phi^{(h)}(A+A';T) \tilde{\Phi}^{(h)}(A';T), \nonumber
\end{eqnarray}
scale out in $D_N <0$, which is certainly satisfied in the closed CDT condition.
Furthermore, we can see the suppression of the IK-type interactions to the surfaces with more than one boundary loops, which are derived from the omitted terms in the dots of the effective action (\ref{eq:effectiveaction}).
By using the definition of the field with arbitrary loop number $b$, Eq.(\ref{eq:open}), we add the higher order terms of IK-type interaction,
\begin{eqnarray}
\sum_{b=2}^{\infty} a^{2b-2} G_{\rm{st}}^{b \over 2} \int_0^{\infty} dA' \int_0^{\infty}dL_1 \cdots \int_0^{\infty}dL_b {A' \over L_1 \cdots L_b} \Omega^{(0|b)}(A+A'|L_1,\cdots ,L_b;T) \tilde{\Omega}^{(0|b)}(A'|L_1,\cdots ,L_b;T), \nonumber
\end{eqnarray}
after the leading term of (\ref{eq:FPH5}).
They scale out for $b \ge 2$.
This restriction of the IK-type is desirable because if it were not for this suppression, we had infinite series of IK-type interaction increasing any number of boundary loops.
Then, the IK-type interactions with preserving loop number and with adding loop number only by one were relatively suppressed in a pile of interactions.
The existence of these higher order interactions tended to increase the boundary loop number, then the closed surfaces were not stable any more.
}.
The surface field theory based on CDT with the IK-type interactions is realized in the continuum limit when all terms with the explicit power indices of the scaling parameter $a$ in the above FP Hamiltonian become positive.
As mentioned in the previous section, we divide interactions into three groups, then we estimate every term in each group in turn (see Fig.\ref{fig:Hamiltonian}).

At first, let us focus on the first group of the interactions, or the deformation of a closed surface, the terms from (\ref{eq:FPH1}) to (\ref{eq:FPH7}).
In order to take the infinitesimal limit of discrete stochastic time interval for the continuum one, we obtain the first condition for Eq.(\ref{eq:stochastic-time}),
\begin{eqnarray}
\label{eq:restriction0}
D>8.
\end{eqnarray}
In the closed surface CDT model, the interactions of (\ref{eq:FPH1}), (\ref{eq:FPH2}), (\ref{eq:FPH4}) and (\ref{eq:FPH6}) have already estimated in ref.\cite{Kaw3}.
Only the IK-type interaction, (\ref{eq:FPH4}), has remained with the scaling $a^0$.
The propagation in the same time slice, (\ref{eq:FPH1}), is forbidden from the viewpoint of the time-foliation structure.
Thus the second condition on $D$ is
\begin{eqnarray}
\label{eq:restriction1}
D<10.
\end{eqnarray}
This condition became $D<4$, discrepantly from Eq.(\ref{eq:restriction0}), if we did not have the CDT breaking linear and cubic terms in $S_1$ of Eq.({\ref{eq:effectiveaction}), which enhanced the scaling power of the propagation term, (\ref{eq:FPH1}).
For the third restriction, the prohibition of the merging interaction, (\ref{eq:FPH6}), by the causality, also provides the inequality, 
\begin{eqnarray}
\label{eq:restriction2}
D_N<-D+1.
\end{eqnarray}
The fourth condition is for the purpose of the suppression of the handle-adding interaction, (\ref{eq:FPH2}),  
\begin{eqnarray}
\label{eq:restriction3}
D_N<-{1 \over 2}D,
\end{eqnarray}
which is always satisfied if the above three inequalities are realized (see Fig.\ref{fig:region}).
Certainly this interaction should be prohibited because it is, in substance, the merging interaction, not of the two distinct surfaces but of two distant parts of an identical surface field. 
The CDT model of the closed surface is realized under the $D_N$-$D$ conditions of (\ref{eq:restriction0}), (\ref{eq:restriction1}) and (\ref{eq:restriction2}), which we will call ``closed CDT condition''.

Then, we have novel interactions of the closed surface, (\ref{eq:FPH3}), (\ref{eq:FPH5}) and (\ref{eq:FPH7}).
The terms (\ref{eq:FPH5}) and (\ref{eq:FPH7}) express another IK-type and another merging interactions, respectively, with an open surface instead of a closed surface.
While the IK-type (\ref{eq:FPH5}) remains as well as (\ref{eq:FPH4}), the merging  (\ref{eq:FPH7}) scales out  with the same power as (\ref{eq:FPH6}). 
It seems natural because the difference is whether the surfaces of interacting target have a boundary loop at the irrelevant position to this interaction.
Baby loop-adding interaction, (\ref{eq:FPH3}), scales out because the condition
\begin{eqnarray}
\label{eq:restriction4}
D_N < -2D+12
\end{eqnarray}
is always satisfied under the closed CDT condition.
This disappearance prevents from the collapse of surfaces by increasing holes hence it assures the stability of the closed surface.

The second group of the interactions, terms from (\ref{eq:FPH8}) to (\ref{eq:FPH14}), expresses the open surface deformation caused at points on surfaces.
We notice that they are similar interactions with exactly the same scaling as those of closed surfaces, from (\ref{eq:FPH1}) to (\ref{eq:FPH7}), respectively, when we decide the scaling of cylinder field as Eq.(\ref{eq:cylinder1}).
We have no new information from them and it is natural because the difference is again the existence or non-existence of a boundary loop at the irrelevant place on the original surface.

The third group, (\ref{eq:FPH15}), (\ref{eq:FPH16}), (\ref{eq:FPH17}), and (\ref{eq:FPH18}), describes the open surface deformation happening at the boundary loops.
It includes the IK-type interaction on the boundary loop, (\ref{eq:FPH17}), without which we have no novelty in the open surface CDT model relative to the closed one.
The propagation of a loop in a time slice, (\ref{eq:FPH15}), happens to take the same scaling as those of the baby loop-adding interactions, (\ref{eq:FPH3}) and (\ref{eq:FPH10}).
It scales out in the condition of Eq.(\ref{eq:restriction4}) as it is expected for the time-foliation structure.
Then the loop-splitting interaction, (\ref{eq:FPH16}), is suppressed in the scaling by
\begin{eqnarray}
\label{eq:restriction5}
D_N<-D+4, 
\end{eqnarray}
which is naturally satisfied under the closed CDT condition, by Eq.(\ref{eq:restriction2}).
Though it is the splitting interaction of a boundary loop on the open surface, it can be seen as the merging interaction of two marginal parts of the surface, hence it is favorable to be prohibited by the causality as well as the merging interaction.
The fact that the above restriction is weaker than that of merging prohibition means that the stability of the surface, preventing from increasing holes, is more fundamental as it should be.
The last interaction (\ref{eq:FPH18}), or the merging interaction of two surfaces by joining at each boundary loop, scales out with the same condition (\ref{eq:restriction2}) as the ordinary four merging interactions, (\ref{eq:FPH6}), (\ref{eq:FPH7}), (\ref{eq:FPH13}) and (\ref{eq:FPH14}).

After all, any term multiplied explicitly with the power of $a$ in the above FP Hamiltonian ${\cal H}_{\rm FP}$ scales out in the closed CDT condition.
Then, we realize the open-closed surface CDT model which includes the IK-type interactions occurring at a point on the surface as well as the boundary loop, as the only quantum effect. 
\begin{figure}[t]
\begin{center}
\includegraphics [width=80mm, height=62mm]{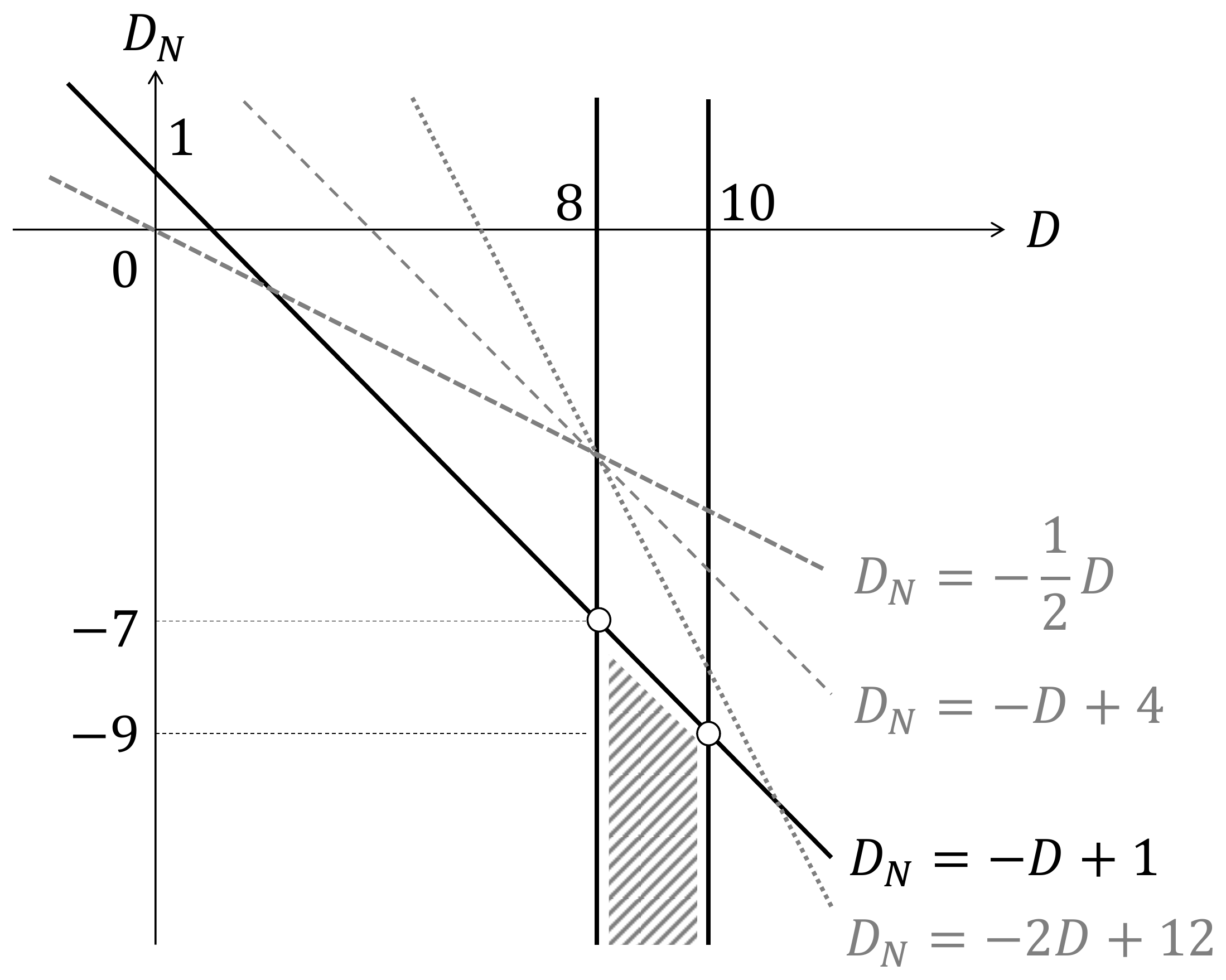}
\caption{On the shaded area in the $D_N$-$D$ space of the scaling dimensions, or the ``closed CDT'' region, we obtain the open-closed CDT model with the IK-type interaction. Whole the area is below three kinds of broken lines. The first condition, $D_N < -{1 \over 2}D$, is the restriction about the scaling out of the handle-adding interaction. The second one, $D_N < -2D+12$, is for the prohibition of the baby loop-adding interaction. The third one, $D_N < -D+4 $, is necessary for avoiding loop-splitting on the open surface. All conditions are naturally satisfied in the closed CDT condition. }
\label{fig:region}
\end{center}
\end{figure}

%
Lastly, let us confirm the possibility to go back to the pure CDT model, which does not leave any interaction including IK-type one in a scaling limit.
If we take the scaling of stochastic time as $d\tau =a^{{1 \over 2}D-4-\varepsilon} \Delta \tau$ with $\varepsilon >0$, differently from Eq.(\ref{eq:stochastic-time}), all the IK-type interactions  (\ref{eq:FPH4}), (\ref{eq:FPH5}), (\ref{eq:FPH11}), (\ref{eq:FPH12}), (\ref{eq:FPH17}) scale out.
Then the conditions of the scaling dimensions to forbid other interactions are modified as follows:
\\
\begin{tabular}{lll}
$\cdot$ Stochastic time becoming continuum $d\tau$ & : & $D>8+2\varepsilon$ \\
$\cdot$ Suppression of merging interactions (\ref{eq:FPH6})(\ref{eq:FPH7})(\ref{eq:FPH13})(\ref{eq:FPH14})(\ref{eq:FPH18}) & : & $D_N < -D+1+\varepsilon$ \\
$\cdot$ Suppression of surface propagation in a time (\ref{eq:FPH1})(\ref{eq:FPH8}) & : & $D<10+2\varepsilon$ \\
$\cdot$ Suppression of handle-adding interactions (\ref{eq:FPH2})(\ref{eq:FPH9}) & : & $D_N < -{1 \over 2}D + \varepsilon$ \\
$\cdot$ Suppression of loop-adding (\ref{eq:FPH3})(\ref{eq:FPH10}) and loop propagation (\ref{eq:FPH15}) & : & $D_N < -2D+12+4\varepsilon$ \\
$\cdot$ Suppression of loop splitting interaction (\ref{eq:FPH16})  & : & $D_N < -D+4+2\varepsilon$  \\
\end{tabular}
\\
Certainly, we derive the pure CDT model by turning on the positive $\varepsilon$, with the area satisfying all the above conditions in the $D_N$-$D$ space, which we obtain by the corresponding shift from the ``closed CDT'' area.
%
\section{Algebraic structure}\label{sec:algebra}
In this section, we go back to the discrete level of the FP Hamiltonian to discuss the algebraic structure in the tensor-matrix model.
We arrange the FP Hailtonian of Eq.(\ref{eq:FPH}) with an additional index of boundary as
\begin{eqnarray}
\label{eq:FPHgenerator}
H_{\rm FP} & = &  \sum_t \left[ {1 \over N^2} \sum_{h=0}^{\infty} \sum_{n=1}^{\infty} n L_t^{(h)} (n-2) \pi _t^{(h)} (n) \right. \nonumber \\
& & +{1 \over N^2} \sum_{\alpha} \sum_{h=0}^{\infty} \sum_{n=1}^{\infty} \sum_{k=1}^{\infty} n K_t^{(h|1) \alpha}(n-2|k) \pi _t^{(h|1) \alpha}(n|k) + \cdots \nonumber \\
& & \left. +{1 \over N} \sum_{\alpha} \sum_{h=0}^{\infty} \sum_{n=1}^{\infty} \sum_{k=1}^{\infty} \lambda_{\rm B}^{\alpha} k J_t^{(h|1) \alpha}(n|k-2) \pi _t^{(h|1) \alpha}(n|k) + \cdots \right],
\end{eqnarray}
where the three kinds of generator, $L_t^{(h)} (n)$, $K_t^{(h|1) \alpha}(n|k)$ and $J_t^{(h|1) \alpha}(n|k)$, concern the deformation of the closed surface, the open surface and the boundary loop of the open surface, respectively. 
The upper index ``$\alpha$'' is assigned to the boundary loop of the disc, similarly to the Chan-Paton factor of the open string in the 2D model.
The open surface attaches its boundary loop on a D-brane with the index ``$\alpha$'', one of the D-branes located at the same position.
In the dots of the second line we hide the sequence of similar terms to the previous one, each of which contains the annihilation operator of a surface with two or more boundary loops and the corresponding generator, 
\begin{eqnarray}
&& {1 \over N^2} \sum_{\alpha, \alpha'} \sum_{h=0}^{\infty} \sum_{n=1}^{\infty} \sum_{k=1}^{\infty} \sum_{k'=1}^{\infty} n K_t^{(h|2) \alpha \alpha' }(n-2|k,k') \pi _t^{(h|2) \alpha \alpha' }(n|k,k') \nonumber \\
&& + {1 \over N^2} \sum_{\alpha, \alpha', \alpha''} \sum_{h=0}^{\infty} \sum_{n=1}^{\infty} \sum_{k=1}^{\infty} \sum_{k'=1}^{\infty} \sum_{k''=1}^{\infty} n K_t^{(h|3) \alpha \alpha' \alpha'' }(n-2|k,k',k'') \pi _t^{(h|3) \alpha \alpha' \alpha'' }(n|k,k',k'') +\cdots. \nonumber
\end{eqnarray}
In the above expression, the indices $\alpha,\alpha',\alpha'',\cdots$ are assigned to the loops with length $k,k',k'',\cdots$, respectively. 
Similarly, dots in the third line of Eq.(\ref{eq:FPHgenerator}) includes generators for the deformation of surfaces with more than one boundary loops.
As the next order, the terms with generators concerning a surface with two boundary loops, are
\begin{eqnarray}
{1 \over N} \sum_{\alpha, \alpha'} \sum_{h=0}^{\infty} \sum_{n=1}^{\infty} \sum_{k=1}^{\infty} \sum_{k'=1}^{\infty} \left\{ \lambda_{\rm B}^{\alpha} k J_t^{(h|2) \overline{\alpha} \alpha'}(n|\overline{k-2},k') + \lambda_{\rm B}^{\alpha'} k' J_t^{(h|2) \alpha \overline{\alpha'}}(n|k,\overline{k'-2}) \right\} \pi _t^{(h|2) \alpha \alpha'}(n|k,k'), \nonumber
\end{eqnarray}
where over-lined variables in the generator specify the boundary loop on which the corresponding deformation occurs.

The explicit form of the first generator, for a closed surface, is
\begin{eqnarray}
\label{eq:generator1}
L_t^{(h)} (n) & \equiv & -N^2 \left[ g \phi _t^{(h)} (n+1) - \phi _t^{(h)} (n+2) + g \phi _t^{(h)} (n+3) + {1 \over N}(n+1) \phi_t^{(h+1)} (n) \right. \nonumber \\
& & + {1 \over N} g_{\rm B} \sum_{\alpha'} \left\{ {1 \over 2} \omega _t^{(h|1) \alpha'} (n+1|2) + {1 \over 4} \omega _t^{(h|1) \alpha'} (n+1|4) \right\} \nonumber \\
& & + \sum_{h'=0}^{\infty} \sum_{n'=1}^{\infty} \phi_t^{(h+h')} (n+n') \tilde{\phi}_t^{(h')} (n') \nonumber \\
& & + {1 \over N} \sum_{h'=0}^{\infty} \sum_{\alpha'} \sum_{n'=1}^{\infty} \sum_{k'=1}^{\infty} {n' \over k'} \omega _t^{(h+h'|1)\alpha'} (n+n'|k') \tilde{\omega}_t^{(h'|1)\alpha'} (n'|k') +\cdots \nonumber \\
& & + {1 \over N^2} \sum_{h'=0}^{\infty} \sum_{n'=1}^{\infty} n' \phi_t^{(h+h')}(n+n') \pi_t^{(h')} (n') \nonumber \\
& & \left. + {1 \over N^2} \sum_{h'=0}^{\infty} \sum_{\alpha'} \sum_{n'=1}^{\infty} \sum_{k'=1}^{\infty} n' \omega_t^{(h+h'|1)\alpha'}(n+n'|k') \pi_t^{(h'|1)\alpha'} (n'|k') +\cdots \right] ,
\end{eqnarray}
where, of course, the dots contain the terms concerning open surfaces with more than one boundary loops.
The generator of the deformation at a point on the surface of a disc, not on the boundary loop, is
\begin{eqnarray}
\label{eq:generator2}
K_t^{(h|1)\alpha} (n|k) & \equiv & -N^2 \left[ g \omega _t^{(h|1)\alpha} (n+1|k) - \omega _t^{(h|1)\alpha} (n+2|k) + g \omega _t^{(h|1)\alpha} (n+3|k) \right. \nonumber \\
& & + {1 \over N}(n+1) \omega_t^{(h+1|1)\alpha} (n|k) \nonumber \\
& & + {1 \over N} g_{\rm B} \sum_{\alpha'} \left\{ {1 \over 2} \omega _t^{(h|1) \alpha \alpha'} (n+1|k,2) + {1 \over 4} \omega _t^{(h|1) \alpha \alpha'} (n+1|k,4) \right\} \nonumber \\
& & + \sum_{h'=0}^{\infty} \sum_{n'=1}^{\infty} \omega_t^{(h+h'|1)\alpha} (n+n'|k) \tilde{\phi}_t^{(h')} (n') \nonumber \\
& & + {1 \over N} \sum_{h'=0}^{\infty} \sum_{\alpha'} \sum_{n'=1}^{\infty} \sum_{k'=1}^{\infty} {n' \over k'} \omega _t^{(h+h'|2)\alpha \alpha'} (n+n'|k,k') \tilde{\omega}_t^{(h'|1)\alpha'} (n'|k') +\cdots \nonumber \\
& & + {1 \over N^2} \sum_{h'=0}^{\infty} \sum_{n'=1}^{\infty} n' \omega_t^{(h+h'|1)\alpha}(n+n'|k) \pi_t^{(h')} (n') \nonumber \\
& & + {1 \over N^2} \sum_{h'=0}^{\infty} \sum_{\alpha'} \sum_{n'=1}^{\infty} \sum_{k'=1}^{\infty} n' \omega_t^{(h+h'|2)\alpha \alpha'}(n+n'|k,k') \pi_t^{(h'|1)\alpha'} (n'|k') \nonumber \\
& & + {1 \over N^2} \sum_{h'=0}^{\infty} \sum_{\alpha' \alpha''} \sum_{n'=1}^{\infty} \sum_{k',k''=1}^{\infty} n' \omega_t^{(h+h'|3)\alpha \alpha' \alpha''}(n+n'|k,k',k'') \pi_t^{(h'|2)\alpha' \alpha''} (n'|k',k'') \nonumber \\
& & \left.  +\cdots \right] ,
\end{eqnarray}
where dots in the fifth line contain the similar terms with higher number of boundary loops.
In the eighth line we write down explicitly the second order terms, following to the previous terms of the lower orders.
The dots in the last line begin with the next order.
We see the first generator, $L_t^{(h)} (n)$, is only the special case with the number of boundary loop zero, $K_t^{(h|0)}(n)$.
The generator concerning the boundary loop of the same disc is
\begin{eqnarray}
\label{eq:generator3}
J_t^{(h|1)\alpha} (n|k) & \equiv & -N \left[ - \omega _t^{(h|1)\alpha} (n|k+2) + g_{\rm B} \left\{ \omega _t^{(h|1)\alpha} (n+1|k+2) + \omega _t^{(h|1)\alpha} (n+1|k+4) \right\} \right. \nonumber \\
& & + {1 \over N} \sum_{k'=0}^k \omega_t^{(h|2)\alpha \alpha} (n|k',k-k') \nonumber \\
& & + \sum_{h'=0}^{\infty} \sum_{n'=1}^{\infty} \sum_{k'=1}^{\infty} \omega_t^{(h+h'|1)\alpha} (n+n'|k+k') \tilde{\omega}_t^{(h'|1)\alpha} (n'|k') +\cdots  \nonumber \\
& & + {1 \over N} \sum_{h'=0}^{\infty} \sum_{n'=1}^{\infty} \sum_{k '=1}^{\infty} k' \omega _t^{(h+h'|1)\alpha} (n+n'|k+k') \pi _t^{(h'|1)\alpha} (n'|k') \nonumber \\
& & + {1 \over N} \sum_{h'=0}^{\infty} \sum_{\alpha'} \sum_{n'=1}^{\infty} \sum_{k'=1}^{\infty} \sum_{k''=1}^{\infty} k'' \omega_t^{(h+h'|2)\alpha \alpha'}(n+n'|k+k'',k') \pi_t^{(h'|2)\alpha \alpha'} (n'|k'',k') \nonumber \\
& & \left. +\cdots \right] ,
\end{eqnarray}
where dots include same meaning as the previous generators.
In the above expression for disc, with the minimum number of loop, we have omitted the over-line to specify the loop. 

First two generators satisfy the following commutation relations: 
\begin{eqnarray}
\label{eq:virasoro1}
\left[ L_t^{(h)} (n) , L_{t'}^{(g)} (m) \right] &=& (n-m) \delta_{t t'} L_t^{(h+g)} (n+m) ,\\
\label{eq:virasoro2}
\left[ K_t^{(h|1)\alpha} (n|k) , K_{t'}^{(g|1)\beta} (m|\ell) \right] &=& (n-m) \delta_{t t'} K_t^{(h+g|2)\alpha \beta} (n+m|k,\ell) ,\\
\label{eq:virasoro3}
\left[ K_t^{(h|1)\alpha} (n|k) , L_{t'}^{(g)} (m) \right] &=& (n-m) \delta_{t t'} K_t^{(h+g|1)\alpha} (n+m|k) .
\end{eqnarray}
They construct the Virasoro-like algebra relating to the surface area variable.
Also from the viewpoint of algebra, the generator $L_t^{(h)} (n)$ is in the same kind as $K_t^{(h|1)\alpha} (n|k)$ with the link shrinking to disappear, $k \rightarrow 0$, differently from the 2D case.
We regard they are the two lowest order members in the infinite series of open surface generators with the order number of boundary loop.
In general, we expect the generators $K_t^{(h|p)\alpha_1 \cdots \alpha_p} (n|k_1,\cdots ,k_p)$, which we can write down easily as the extension of Eqs.(\ref{eq:generator1}) and (\ref{eq:generator2}), for any number of boundary loops $p(=0,1,2,3,\cdots)$, to satisfy the commutation relation
\begin{eqnarray}
\label{eq:virasoro4}
&&\left[ K_t^{(h|p)\alpha_1 \cdots \alpha_p} (n|k_1,\cdots ,k_p) , K_{t'}^{(g|q)\beta_1 \cdots \beta_q} (m|\ell_1,\cdots ,\ell_q) \right] \nonumber \\
&&\hspace{10mm} = (n-m) \delta_{t t'} K_t^{(h+g|p+q)\alpha_1\cdots \alpha_p  \beta_1\cdots \beta_q} (n+m|k_1,\cdots ,k_p,\ell_1,\cdots ,\ell_q) .
\end{eqnarray}
The commutation relation of the third generators is
\begin{eqnarray}
\label{eq:virasoro5}
\left[ J_t^{(h|1)\alpha} (n|k) , J_{t'}^{(g|1 )\beta} (m|\ell ) \right] &=& (k-\ell ) \delta_{t t'}\delta_{\alpha \beta} J_t^{(h+g|1)\alpha} (n+m|k+\ell ) .
\end{eqnarray}
It expresses another Virasoro-like algebraic structure, like the 2D CDT model of string.
We obtain the commutation relation of the first and third generators as well as the one of the second and third generators.
The latter seems certainly the extension of the former as
\begin{eqnarray}
\label{eq:virasoro6}
\left[ J_t^{(h|1)\alpha} (n|k) , L_{t'}^{(g)} (m) \right] &=& n \delta_{t t'} J_t^{(h+g|1) \alpha } (n+m| k), \nonumber \\
\left[ J_t^{(h|1)\alpha} (n|k) , K_{t'}^{(g|1 )\beta} (m|\ell ) \right] &=&  n \delta_{t t'} J_t^{(h+g|2)\overline{\alpha} \beta}(n+m|\overline{k},\ell) - \ell \delta_{t t'}\delta_{\alpha \beta} K_t^{(h+g|1)\alpha} (n+m|k+\ell ) . \nonumber \\
~
\end{eqnarray}
In the latter expression, we recognize the generator for the deformation happening at a link on one boundary loop of the cylinder: 
\begin{eqnarray}
\label{eq:generator4}
& & \hspace{-10mm} J_t^{(h|2)\overline{\alpha} \beta}(n|\overline{k},\ell)  \nonumber \\
& & \equiv -N \left[ - \omega _t^{(h|2)\alpha \beta} (n|k+2, \ell ) \right. \nonumber \\
& & + g_{\rm B} \left\{ \omega _t^{(h|2)\alpha \beta} (n+1|k+2, \ell ) + \omega _t^{(h|2)\alpha \beta } (n+m+1|k+4, \ell ) \right\} \nonumber \\
& & + \delta _{\alpha \beta} \ell \omega _t^{(h+1|1)\alpha} (n|k+\ell ) \nonumber \\
& & + {1 \over N} \sum_{k'=0}^k \omega_t^{(h|3)\alpha \alpha \beta} (n|k',k-k',\ell ) \nonumber \\
& & + \sum_{h'=0}^{\infty} \sum_{n'=1}^{\infty} \sum_{k'=1}^{\infty} \omega_t^{(h+h'|2)\alpha \beta } (n+n'|k+k',\ell ) \tilde{\omega}_t^{(h'|1)\alpha} (n'|k') +\cdots  \nonumber \\
& & + {1 \over N} \sum_{h'=0}^{\infty} \sum_{n'=1}^{\infty} \sum_{k '=1}^{\infty} k' \omega _t^{(h+h'|2)\alpha \beta } (n+n'|k+k',\ell ) \pi _t^{(h'|1)\alpha} (n'|k') \nonumber \\
& & + {1 \over N} \sum_{h'=0}^{\infty} \sum_{\alpha'} \sum_{n'=1}^{\infty} \sum_{k'=1}^{\infty} \sum_{k''=1}^{\infty} k' \omega_t^{(h+h'|3)\alpha \alpha' \beta}(n+n'|k+k',k'', \ell ) \pi_t^{(h'|2)\alpha \alpha'} (n'|k',k'') \nonumber \\
& & \left. +\cdots \right] .
\end{eqnarray}
Notice that it has the same structure with the generator of disc, Eq.(\ref{eq:generator3}), except for one additional term of the fourth line, which certainly should be included for the closure of the algebra (Fig.\ref{fig:int19}).
\begin{figure}[t]
\begin{center} 
\includegraphics [width=90mm, height=25mm]{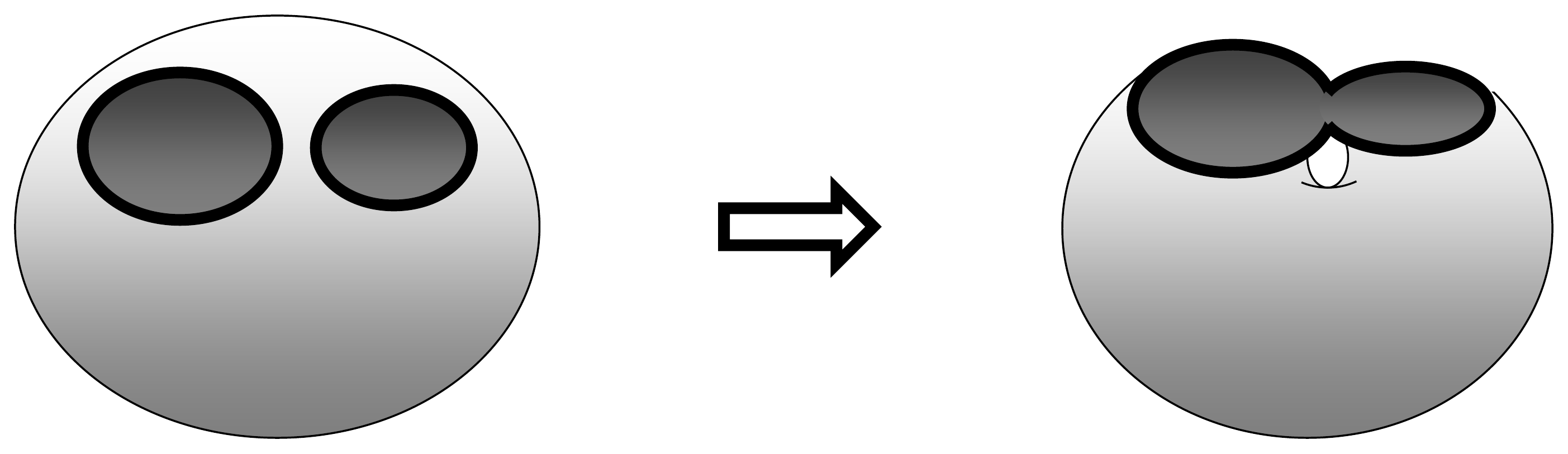}
\caption{Merging interaction of two boundary loops of a cylinder surface field is possible only in the case the two loops belong to the same D-brane, $\alpha = \beta $. The model includes this interaction naturally for the symmetry, or the closure of commutation relations.}
\label{fig:int19}
\end{center}
\end{figure}
This term corresponds to the merging interaction of both boundary loops of the cylinder to generate open surface with one loop and one handle, or the deformation diminishing loop number by one and increasing handle number by one.
Though we have confirmed the algebraic structure for the generators only in the lowest order of the loop number, in any order we expect the commutation relations to close within the infinite series of generators of two types.
\section{Conclusions}
We have constructed 3D CDT open-closed surface field theory from the tensor-matrix model in the same way as the description of 2D CDT string field theory led from matrix-vector model describing the loop gas model.
The effective action includes the interaction of the one-step propagators deriving the CDT time-foliation structure, with extra CDT breaking terms.
The stochastic quantization method is utilized to add the quantum correction to the simple propagation model of exact CDT.
The stochastic time is interpreted as the growth of quantum effect, not the geodesic distance.
Differently from the 2D string generalized CDT model, splitting interaction can not be contained in the stochastic quantization procedure therefore the IK-type interaction is only the possibility of the quantum effect.
Through the double scaling limit, the CDT model with additional IK-type interaction is realized.
When we assume the whole IK-type interactions remain, all uninvited interactions for CDT, contained at the discrete level, scale out.
The condition for the scaling dimensions $D_N$ and $D$ to realize this open-closed surface model is exactly same as the one for the case of the closed CDT model.
We find the open surface interacts in the same manner with the same scaling as the closed surface, as long as the interaction occurs on the surface, not at the boundary.
A common property with the 2D CDT model is that all the merging interactions scale by the factor $a^{-D-D_N+1}$ higher than the remaining interactions, including the IK-type interactions.
The interactions coming about on the boundary loop are similar to the 2D string CDT interactions.
Although, the loop splitting interaction, which is permitted in the 2D original string model, is forbidden consistently in the interpretation that it is seen as the inconvenient merging interaction of the two distant points from the viewpoint of the surface.
The permission of the splitting interaction may be the special circumstances in the string field 2D model.
It is conjectured that in four and higher dimensional CDT, the situation is not changed that we may realize the CDT space field theory which contains only propagation and IK-type interaction.
Furthermore, it may be possible to make the tensor model contain space field, surface field and string field by providing various kinds of tensor field. 
This 3D CDT model and the 2D CDT model are expected to be contained as the subspace field theory in the D-branes of the CDT model in higher dimensional space-time, except for the string splitting interactions.

Then, we have investigated the algebraic structure of this model.
We have infinite series of generators, which are classified into two types.
Though our investigation is limited only for the lowest orders, we are sure the exact closure of the generators in their commutators.
It is the advantage of the 3D model to the 2D model, where the commutators concerning the generator of open string deformation at edges was not closed within the generators, but left the terms multiplied by some open string fields explicitly.
It was not an algebra in the precise sense, in the 2D model, but was understood as the consistency condition for the constraints.
Meanwhile, in our 3D model, from the viewpoint of the commutation relation, as well as the contents of the generators, the closed surface should be treated as the same kind as the open surface with boundary number zero. 
The symmetry related to the algebra may guarantee that the interactions in the discrete level are consistent as the surface field model.
On the other hand, in the continuum limit, all merging interactions with annihilation operator scale out, hence all generators become to commute with each other.
The 3D CDT surface field model realization from the discrete tensor-matrix model, through the continuum limit, is accompanied with the breakdown of the symmetry concerning algebraic structure of the above one to the simpler one. 

\section*{Acknowledegments}

The author thanks N. Nakazawa for an earlier collaboration.

%

\end{document}